\documentclass{article}

\PassOptionsToPackage{compress,numbers}{natbib}

\usepackage[preprint]{neurips_2019}

\usepackage[utf8]{inputenc} 
\usepackage[T1]{fontenc}    
\usepackage{hyperref}       
\usepackage{url}            
\usepackage{booktabs}       
\usepackage{amsfonts}       
\usepackage{nicefrac}       
\usepackage{microtype}      

\usepackage{graphicx}
\usepackage{dsfont}
\usepackage{wrapfig}

\usepackage{mathtools}

\newcommand{\timeavg}[1]{\left\langle #1 \right\rangle_t}
\newcommand\blfootnote[1]{\begingroup
 \renewcommand\thefootnote{}\footnote{#1}
  \addtocounter{footnote}{-1}
  \endgroup}

\title{Unsupervised Discovery of Temporal Structure in Noisy Data with Dynamical Components Analysis}

\author{
David G. Clark$^{*,1,2}$ \:\:\: Jesse A.~Livezey$^{*,2,3}$ \:\:\: Kristofer E. Bouchard$^{2,3,4}$ \\
\texttt{dgc2138@cumc.columbia.edu} \:\:\:
\texttt{jlivezey@lbl.gov} \:\:\:
\texttt{kebouchard@lbl.gov} \\
$^*$Equal contribution. \\
$^1$Center for Theoretical Neuroscience, Columbia University \\
$^2$Biological Systems and Engineering Division, Lawrence Berkeley National Laboratory \\
$^3$Redwood Center for Theoretical Neuroscience, University of California, Berkeley \\
$^4$Helen Wills Neuroscience Institute, University of California, Berkeley}

\begin{document}

\maketitle

\begin{abstract}

Linear dimensionality reduction methods are commonly used to extract low-dimensional structure from high-dimensional data. However, popular methods disregard temporal structure, rendering them prone to extracting noise rather than meaningful dynamics when applied to time series data. At the same time, many successful unsupervised learning methods for temporal, sequential and spatial data extract features which are predictive of their surrounding context. Combining these approaches, we introduce Dynamical Components Analysis (DCA), a linear dimensionality reduction method which discovers a subspace of high-dimensional time series data with maximal predictive information, defined as the mutual information between the past and future. We test DCA on synthetic examples and demonstrate its superior ability to extract dynamical structure compared to commonly used linear methods. We also apply DCA to several real-world datasets, showing that the dimensions extracted by DCA are more useful than those extracted by other methods for predicting future states and decoding auxiliary variables. Overall, DCA robustly extracts dynamical structure in noisy, high-dimensional data while retaining the computational efficiency and geometric interpretability of linear dimensionality reduction methods. 
\end{abstract}

\section{Introduction}Extracting meaningful structure from noisy, high-dimensional data in an unsupervised manner is a fundamental problem in many domains including neuroscience, physics, econometrics and climatology. In the case of time series data, e.g., the spiking activity of a network of neurons or the time-varying prices of many stocks, one often wishes to extract features which capture the dynamics underlying the system which generated the data. Such dynamics are often expected to be low-dimensional, reflecting the fact that the system has fewer effective degrees of freedom than observed variables. For instance, in neuroscience, recordings of 100s of neurons during simple stimuli or behaviors generally contain only $\sim$10 relevant dimensions~\citep{gao2015simplicity}. In such cases, dimensionality reduction methods may be used to uncover the low-dimensional dynamical structure.\blfootnote{DCA code is available at: \url{https://github.com/BouchardLab/DynamicalComponentsAnalysis}}

Linear dimensionality reduction methods are popular since they are computationally efficient, often reducing to generalized eigenvalue or simple optimization problems, and geometrically interpretable, since the high- and low-dimensional variables are related by a simple change of basis~\citep{cunningham2015linear}. Analyzing the new basis can provide insight into the relationship between the high- and low-dimensional variables~\citep{mahoney2009cur}. However, many popular linear methods including Principal Components Analysis, Factor Analysis and Independent Components Analysis disregard temporal structure, treating data at different time steps as independent samples from a static distribution. Thus, these methods do not recover dynamical structure unless it happens to be associated with the static structure targeted by the chosen method.

On the other hand, several sophisticated unsupervised learning methods for temporal, sequential and spatial data have recently been proposed, many of them rooted in \textit{prediction}. These prediction-based methods extract features which are predictive of the future (or surrounding sequential or spatial context)~\citep{mikolov2013efficient, doersch2015unsupervised, kim2016character, marzen2017nearly, mcallester2018information, oord2018representation}. Predictive features form useful representations since they are generally linked to the dynamics, computation or other latent structure of the system which generated the data. Predictive features are also of interest to organisms, which must make internal estimates of the future of the world in order to guide behavior and compensate for latencies in sensory processing \cite{bouchard2016auditory}. These ideas have been formalized mathematically~\citep{tishby2000information, bialek2001predictability} and tested experimentally~\citep{palmer2015predictive}.

We introduce Dynamical Components Analysis (DCA), a novel method which combines the computational efficiency and ease of interpretation of linear dimensionality reduction methods with the temporal structure-discovery power of prediction-based methods. Specifically, DCA discovers a subspace of high-dimensional time series data with maximal predictive information, defined as the mutual information between the past and future~\citep{bialek2001predictability}. To make the predictive information \text{differentiable} and accurately estimable, we employ a Gaussian approximation of the data, however we show that maximizing this approximation can yield near-optima of the full information-theoretic objective. We compare and contrast DCA with several existing methods, including Principal Components Analysis and Slow Feature Analysis, and demonstrate the superior ability of DCA to extract dynamical structure in synthetic data. We apply DCA to several real-world datasets including neural population activity, multi-city weather data and human kinematics. In all cases, we show that DCA outperforms commonly used linear dimensionality reduction methods at predicting future states and decoding auxiliary variables. Altogether, our results establish that DCA is an efficient and robust linear method for extracting dynamical structure embedded in noisy, high-dimensional time series. 

\section{Dynamical Components Analysis}

\subsection{Motivation}
\label{subsec:motivation}

\begin{figure}[htbp!]
	\centering
	\includegraphics[width=\textwidth]{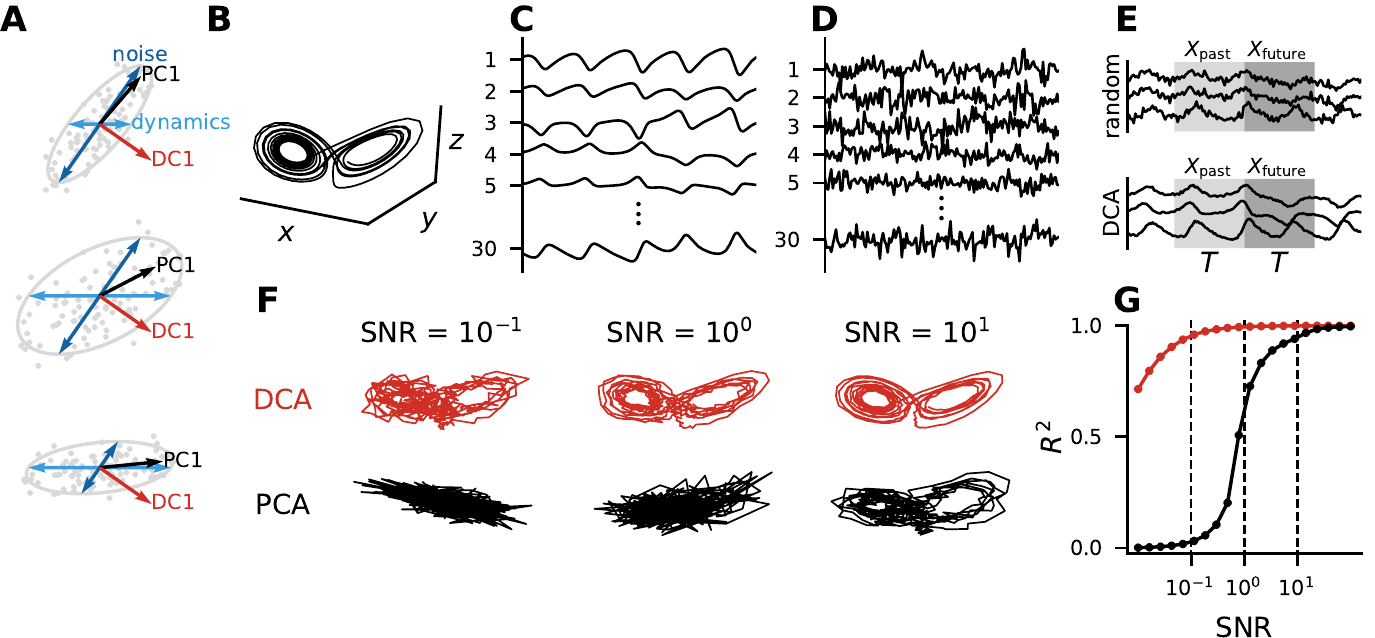}
	\caption{\textbf{DCA finds dynamics rather than variance.} \textbf{(A)} Schematic of unit vectors found by PCA and DCA for three relative levels of dynamics and noise. The dimension of greatest variance, found by PCA, contains contributions from both sources while the dimension found by DCA is orthogonal to the noise. \textbf{(B)} Lorenz attractor in the chaotic regime. \textbf{(C)} Random orthogonal embedding of the Lorenz attractor into 30-dimensional space. \textbf{(D)} Embedded Lorenz attractor with spatially-structured white noise. \textbf{(E)} Random three-dimensional projection (top) and DCA projection (bottom) of the embedded Lorenz attractor. \textbf{(F)} Reconstructions of the Lorenz attractor given the three-dimensional projections found by DCA and PCA. \textbf{(G)} Lorenz reconstruction performance ($R^2$) as a function of the SNR for both methods. See Appendix B for details of the noisy Lorenz embedding.}
	\label{fig:lorenz}
\end{figure}

Dimensionality reduction methods that do not take time into account will miss dynamical structure that is not associated with the static structure targeted by the chosen method. We demonstrate this concretely in the context of Principal Components Analysis (PCA), whose static structure of interest is \textit{variance} \citep{pearson1901liii, hotelling1933analysis}. Variance arises in time series due to both dynamics and noise, and the dimensions of greatest variance, found by PCA, contain contributions from both sources in general. Thus, PCA is prone to extracting spatially structured noise rather than dynamics if the noise variance dominates, or is comparable to, the dynamics variance (Fig.~\ref{fig:lorenz}A). We note that for applications in which generic shared variability due to both dynamics and spatially structured noise is of interest, static methods are well-suited.

To further illustrate this failure mode of PCA, suppose we embed a low-dimensional dynamical system, e.g., a Lorenz attractor, in a higher-dimensional space via a random embedding (Fig.~\ref{fig:lorenz}B,C). We then add spatially anisotropic Gaussian white noise (Fig.~\ref{fig:lorenz}D). We define a signal-to-noise ratio (SNR) given by the ratio of the variances of the first principal components of the dynamics and noise. When the SNR is small, the noise variance dominates the dynamics variance and PCA primarily extracts noise, missing the dynamics. Only when the SNR becomes large does PCA extract dynamical structure (Fig.~\ref{fig:lorenz}F,G, black). Rather than maximizing variance, DCA finds a projection which maximizes the mutual information between past and future windows of length $T$ (Fig.~\ref{fig:lorenz}E). As we will show, this mutual information is maximized precisely when the projected time series contains as much dynamical structure, and as little noise, as possible. As a result, DCA extracts dynamical structure even for small SNR values, and consistently outperforms PCA in terms of dynamics reconstruction performance as the SNR grows (Fig~\ref{fig:lorenz}F,G, red).

\subsection{Predictive information as an objective function}

The goal of DCA is to extract a subspace with maximal dynamical structure. One fundamental characteristic of dynamics is predictability: in a system with dynamics, future uncertainty is reduced by knowledge of the past. This reduction in future uncertainty may be quantified using information theory. In particular, if we equate uncertainty with entropy, this reduction in future uncertainty is the mutual information between the past and future. This quantity was termed predictive information by~\citet{bialek2001predictability}. Formally, consider a discrete time series ${X = \{x_t\}}$, ${x_t \in \mathbb{R}^n}$, with a stationary (time translation-invariant) probability distribution $P(X)$. Let $X_\text{past}$ and $X_\text{future}$ denote consecutive length-$T$ windows of $X$, i.e., ${X_\text{past} = \left( x_{-T+1}, \ldots, x_{0} \right) }$ and ${X_\text{future} = \left(x_1, \ldots, x_{T} \right) }$. Then, the predictive information $I^\text{pred}_T(X)$ is defined as
\begin{equation}
\begin{split}
    I^\text{pred}_T(X)
    &= H\left( X_\text{future} \right) - H\left( X_\text{future} | X_\text{past} \right) \\
    &= H\left( X_\text{past} \right) + H\left( X_\text{future} \right) - H\left( X_\text{past} , X_\text{future}\right) \\
    &= 2H_X(T) - H_X(2T)
\end{split}
\end{equation}
where $H_X(T)$ is the entropy of any length-$T$ window of $X$, which is well-defined by virtue of the stationarity of $X$. Unlike entropy and related measures such as Kolmogorov complexity \citep{li2013introduction}, predictive information is minimized, not maximized, by serially independent time series (white noise). This is because predictive information captures the sub-extensive component of the entropy of $X$. Specifically, if the data points that comprise $X$ are mutually independent, then $H_X(\alpha T) = \alpha H_X(T)$ for all $\alpha$ and $T$, meaning that the entropy is perfectly extensive. On the other hand, if $X$ has temporal structure, then $H_X(\alpha T) < \alpha H_X(T)$ and the entropy has a sub-extensive component given by $\alpha H_X(T) - H_X(\alpha T) > 0$. Upon setting $\alpha = 2$, this sub-extensive component is the predictive information.

Beyond simply being able to detect the presence of temporal structure in time series, predictive information discriminates between different types of structure. For example, consider two discrete-time Gaussian processes with autocovariance functions ${f_1(\Delta t) = \text{exp}\left(- | \Delta t / \tau | \right)}$ and ${f_2(\Delta t) = \text{exp}\left(-\Delta t^2 / \tau^2\right)}$. For $\tau \gg 1$, the predictive information in these time series saturates as ${T \rightarrow \infty}$ to $c_1 \log \frac{\tau}{2}$ and $c_2 \tau^4$, respectively, where $c_1$ and $c_2$ are constants of order unity (see Appendix D for derivation). The disparity in the predictive information of these time series corresponds to differences in their underlying dynamics. In particular, $f_1(\Delta t)$ describes Markovian dynamics, leading to small predictive information, whereas $f_2(\Delta t)$ describes longer-timescale dependencies, leading to large predictive information. Finally, as discussed by \citet{bialek2001predictability}, the predictive information of many time series diverges with $T$. In these cases, different scaling behaviors of the predictive information correspond to different classes of time series. For one-dimensional time series, it was demonstrated that the divergent predictive information provides a unique complexity measure given simple requirements \citep{bialek2001predictability}.

\subsection{The DCA method}

DCA takes as input samples $x_t \in \mathbb{R}^n$ of a discrete time series $X$, as well as a target dimensionality $d \leq n$, and outputs a projection matrix ${V \in \mathbb{R}^{n \times d}}$ such that the projected data $y_t = V^T x_t$ maximize an empirical estimate of $I^\text{pred}_T(Y)$. In certain cases of theoretical interest, $P(X)$ is known and $I^\text{pred}_T(Y)$ may be computed exactly for a given projection $V$. Systems for which this is possible include linear dynamical systems with Gaussian noise and Gaussian processes more broadly. In practice, however, we must estimate of $I^\text{pred}_T(Y)$ from finitely many samples. Directly estimating mutual information from multidimensional data with continuous support is possible, and popular nonparametric methods include those based on binning \citep{strong1998entropy, paninski2003estimation}, kernel density estimation \citep{kolchinsky2017estimating} and $k$-nearest neighbor ($k$NN) statistics \citep{kraskov2004estimating}. However, many of these nonparametric methods are not differentiable (e.g., $k$NN-based methods involve counting data points), complicating optimization. Moreover, these methods are typically sensitive to the choice of hyperparameters \citep{zeng2018jackknife} and suffer from the curse of dimensionality, requiring prohibitively many samples for accurate results \citep{mcallester2018formal}.

To circumvent these challenges, we assume that $X$ is a stationary (discrete-time) Gaussian process. It then follows that $Y$ is stationary and Gaussian since $Y$ is a linear projection of $X$. Under this assumption, $I^\text{pred}_T(Y)$ may be computed from the second-order statistics of $Y$, which may in turn be computed from the second-order statistics of $X$ given $V$. Crucially, this estimate of $I^\text{pred}_T(Y)$ is differentiable in $V$. Toward expressing $I^\text{pred}_T(Y)$ in terms of $V$, we define $\Sigma_{T}(X)$, the spatiotemporal covariance matrix of $X$ which encodes all second-order statistics of $X$ across $T$ time steps. Assuming that $\left\langle x_t \right\rangle_t = 0$, we have
\begin{equation}
    \Sigma_{T}(X)
    = \begin{pmatrix}
    C_0 & C_1 & \ldots & C_{T-1} \\
    C_1^T & C_0 & \ldots & C_{T - 2} \\
    \vdots & \vdots & \ddots & \vdots\\
    C_{T-1}^T & C_{T-2}^T & \ldots & C_{0} \\
    \end{pmatrix} \:\: \text{where} \:\:\: C_{\Delta t} = \left\langle x_tx_{t + \Delta t}^T \right\rangle_t.
    \label{eq:saptiotemporal}
\end{equation}

Then, the spatiotemporal covariance matrix of $Y$, $\Sigma_{T}(Y)$, is given by sending $C_{\Delta t} \rightarrow V^T C_{\Delta t} V$ in $\Sigma_{T}(X)$. Finally, $I^\text{pred}_T(Y)$ is given by
\begin{equation}
I^\text{pred}_T(Y) = 2H_Y(T) - H_Y(2T)
= \log|\Sigma_T(Y)| - \frac{1}{2}\log|\Sigma_{2T}(Y)|.
\label{eq:pidef}
\end{equation}
To run DCA on data, we first compute the $2T$ cross-covariance matrices $C_0, \ldots, C_{2T - 1}$, then maximize the expression for $I^\text{pred}_T(Y)$ of Eq.~\ref{eq:pidef} with respect to $V$ (see Appendix A for implementation details). Note that $I^\text{pred}_T(Y)$ is invariant under invertible linear transformations of the columns of $V$. Thus, DCA finds a subspace as opposed to an ordered sequence of one-dimensional projections. 

\begin{figure}[htbp!]
	\centering
	\includegraphics[width=5.5in]{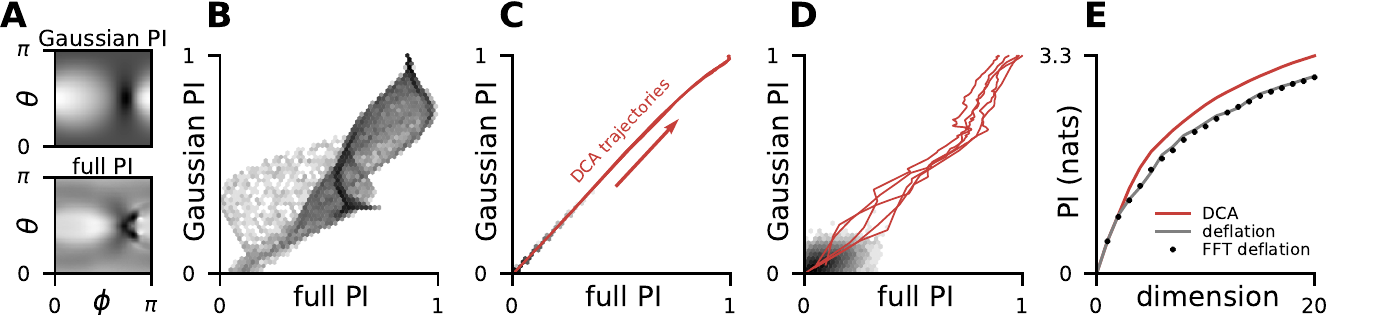}
	\caption{\textbf{Comparison of Gaussian vs. full predictive information estimates (A--D) and the frequency-domain method (E).}  \textbf{(A)} Predictive information of one-dimensional projections of the three-dimensional Lorenz attractor as a function of the spherical coordinates ${(\theta, \phi)}$ of the projection using Gaussian and full ($k$NN) estimates. (A--D) all consider DCA with $T = 1$. \textbf{(B)} Histogram of the Gaussian and full estimates of predictive information from (A). \textbf{(C)} Histogram of the Gaussian and full estimates of predictive information of random one-dimensional projections of the 30-dimensional noisy Lorenz embedding of Fig.~\ref{fig:lorenz}. Red trajectories correspond to five different runs of DCA. \textbf{(D)} Same as (C) but for three-dimensional projections of the same system. \textbf{(E)} Gaussian predictive information of subspaces found by different implementations of DCA when run on 109-dimensional motor cortical data (see Section \ref{sec:results}). ``DCA'' directly optimizes Eq.~\ref{eq:pidef}, ``deflation'' optimizes Eq.~\ref{eq:pidef} to find one-dimensional projections in a deflational fashion and ``FFT deflation'' uses the frequency-domain method of computing Gaussian predictive information in a deflational fashion. $T = 5$ is used in all three cases.}
	\label{fig:lorenz_nonlinear}
\end{figure}

Of course, real data violate the assumptions of both stationarity and Gaussianity. Note that stationarity is a fundamental conceptual assumption of our method in the sense that predictive information is defined only for stationary processes, for which the entropy as a function of window length is well-defined. Nonetheless, extensions of DCA which take nonstationarity into account are possible (see Discussion). On the other hand, the Gaussian assumption makes optimization tractable, but is not required in theory. Note, however, that the Gaussian assumption is acceptable so long as the optima of the Gaussian objective are also near-optima of the full information-theoretic objective. This is a much weaker condition than agreement between the Gaussian and full objectives over all possible $V$. To probe whether the weak condition might hold in practice, we compared the Gaussian estimate of predictive information to a direct estimate obtained using the nonparametric $k$NN estimator of \citet{kraskov2004estimating} for projections of non-Gaussian synthetic data. We refer to these two estimates of predictive information as the ``Gaussian'' and ``full'' estimates, respectively. For random one-dimensional projections of the three-dimensional Lorenz attractor, the Gaussian and full predictive information estimates are positively correlated, but show a complex, non-monotonic relationship (Fig.~\ref{fig:lorenz_nonlinear}A,B). However, for one-dimensional projections of the 30-dimensional noisy Lorenz embedding of Fig.~\ref{fig:lorenz}, we observe tight agreement between the two estimates for random projections (Fig.~\ref{fig:lorenz_nonlinear}C, gray histogram). Running DCA, which by definition increases the Gaussian estimate of predictive information, also increases the full estimate (Fig.~\ref{fig:lorenz_nonlinear}C, red trajectories). When we consider three-dimensional projections of the same system, random projections no longer efficiently sample the full range of predictive information, but running DCA nevertheless increases both the Gaussian and full estimates (Fig.~\ref{fig:lorenz_nonlinear}D, trajectories). These results suggest that DCA finds good optima of the full, information-theoretic loss surface in this synthetic system despite only taking second-order statistics into account.

For a one-dimensional Gaussian time series $Y$, it is also possible to compute the predictive information using the Fourier transform of $Y$ \cite{li1996model}. In particular, when the asymptotic predictive information $I^\text{pred}_{T \rightarrow \infty}(Y)$ is finite, we have $I^\text{pred}_{T \rightarrow \infty}(Y) = \sum_{k=1}^{\infty} k b_k^2$ where $\{b_k\}$ are the so-called \textit{cepstrum coefficients} of $Y$, which are related to the Fourier transform of $Y$ (see Appendix C). When the Fourier transform of $Y$ is estimated for length-$2T$ windows in conjunction with a window function, this method computes a regularized estimate of $I^\text{pred}_T(Y)$. We call this the ``frequency-domain'' method of computing Gaussian predictive information (in contrast the ``time-domain'' method of Eq.~\ref{eq:pidef}). Like the time-domain method, the frequency-domain method is differentiable in $V$. Its primary advantage lies in leveraging the fast Fourier transform (FFT), which allows DCA to be run with much larger $T$ than would be feasible using the time-domain method which requires computing the log-determinant of a $T$-by-$T$ matrix, an $\mathcal{O}\left(T^3\right)$ operation. By contrast, the FFT is $\mathcal{O}\left( T \log T\right)$. However, the frequency-domain method is limited to finding one-dimensional projections. To find a multidimensional projection, one can greedily find one-dimensional projections and iteratively project them out of of the problem, a technique called deflation. However, deflation is not guaranteed to find local optima of the DCA objective since correlations between the projected variables are ignored (Fig.~\ref{fig:lorenz_nonlinear}E). For this reason, we use the time-domain implementation of DCA unless stated otherwise.

\section{Related work}
\label{section:related}

\begin{figure}
\begin{center}
\includegraphics[width=5.5in]{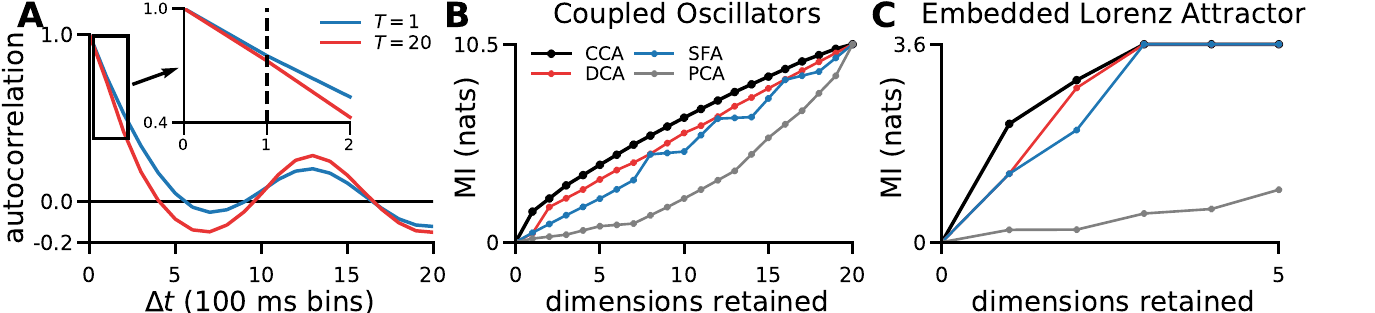}
\end{center}
\caption{\textbf{Comparison of DCA with other methods.} \textbf{(A)} Autocorrelation functions of one-dimensional DCA projections of motor cortical data (see Section \ref{sec:results}) for $T = 1$, in which case DCA is equivalent to SFA, and $T = 20$. While the one-time step autocorrelation is larger for the $T = 1$ projection (inset), the $T = 20$ projection exhibits stronger oscillations apparent at longer timescales. \textbf{(B)} Performance of DCA, SFA, PCA and CCA at capturing past-future mutual information, $I\left(U^T x_t ; V^T x_{t+\Delta t}\right)$, where $U =V$ for DCA, SFA and PCA and $U \neq V$ for CCA. Following \citet{tegmark2019optimal}, $x_t$ comprises the position and momentum variables of 10 coupled oscillators and $\Delta t = 10$. \textbf{(C)} Same as (B), but using the 30-dimensional noisy Lorenz embedding of Fig.~\ref{fig:lorenz} with ${\Delta t = 2}$.}
\label{fig:methods_comparison}
\end{figure}

Though less common than static methods, linear dimensionality reduction methods which take time into account, like DCA, are sometimes used. One popular method is Slow Feature Analysis (SFA), which we examine in some depth due to its resemblance to DCA~\citep{wiskott2002slow, bethge2007unsupervised}. Given a discrete time sereis $X$, where $x_t \in \mathbb{R}^n$, SFA finds projected variables $y_t = V^T x_t \in \mathbb{R}^d$ that have unit variance, mutually uncorrelated components and minimal mean-squared time derivatives. For a discrete one-dimensional time series with unit variance, minimizing the mean-squared time derivative is equivalent to maximizing the one-time step autocorrelation. Thus, SFA may be formulated as
\begin{equation}
     \text{maximize} \:\:\: \text{tr}\left( V^T C_1^\text{sym} V \right) \: \text{subject to} \:\:  V^T C_0 V  = I
\end{equation}
where ${V \in \mathbb{R}^{n \times d}}$, ${C_0 = \timeavg{x_t x_t^T}}$, ${C_1 = \timeavg{x_t x_{t+1}^T}}$ and ${C_1^\text{sym} = \frac{1}{2}\left(C_1 + C_1^T\right)}$. We assume that $X$ has been temporally oversampled so that the one-time step autocorrelation of any one-dimensional projection is positive, which is equivalent to assuming that $C_1^\text{sym}$ is positive-definite (see Appendix E for explanation). SFA is naturally compared to the $T = 1$ case of DCA. For one-dimensional projections (${d = 1}$), the solutions of SFA and DCA coincide, since mutual information is monotonically related to correlation for Gaussian variables in the positive-correlation regime. For higher-dimensional projections ($d > 1$), the comparison becomes more subtle. SFA is solved by making the whitening transformation ${\tilde{V} = C_0^{1/2} V}$ and letting $\tilde{V}$ be the top-$d$ orthonormal eigenvectors of ${M_\text{SFA} = C_0^{-1/2}C_1^\text{sym}  C_0^{-1/2}}$. To understand the solution to DCA, it is helpful to consider the relaxed problem of maximizing ${I(U^T x_t ; V^T x_{t+1})}$ where $U$ need not equal $V$. The relaxed problem is solved by performing Canonical Correlation Analysis (CCA) on $x_t$ and $x_{t+1}$, which entails making the whitening transformations ${\tilde{U} = C_0^{1/2}U}$, ${\tilde{V} = C_0^{1/2}V}$ and letting ${\tilde{U}}$ and $\tilde{V}$ be the top-$d$ left and right singular vectors, respectively, of ${M_\text{CCA} = C_0^{-1/2} C_1 C_0^{-1/2}}$ \cite{borga2001canonical, cunningham2014dimensionality}. If $X$ has time-reversal symmetry, then $C^\text{sym}_1 = C_1$, so $M_\text{SFA} = M_\text{CCA}$ and the projections found by SFA and DCA agree. For time-irreversible processes, $C^\text{sym}_1 \neq C_1$, so $M_\text{SFA} \neq M_\text{CCA}$ and the projections found by SFA and DCA disagree. In particular, the SFA objective has no dependence on the off-diagonal elements of $V^T C_1 V$, while DCA takes these terms into account to maximize ${I\left(V^T x_t ; V^T x_{t+1}\right)}$. Additionally, for non-Markovian processes, SFA and DCA yield different subspaces for $T > 1$ for all $d \geq 1$ since DCA captures longer-timescale dependencies than SFA (Fig.~\ref{fig:methods_comparison}A). In summary, DCA is superior to SFA at capturing past-future mutual information for time-irreversible and/or non-Markovian processes. Note that most real-world systems including biological networks, stock markets and out-of-equilibrium physical systems are time-irreversible. Moreover, real-world systems are generally non-Markovian. Thus, when capturing past-future mutual information is of interest, DCA is superior to SFA for most realistic applications.

With regard to the relaxed problem solved by CCA, \citet{tegmark2019optimal} has suggested that, for time-irreversible processes $X$, the maximum of ${I\left(U^T x_t ; V^T x_{t+1}\right)}$ can be significantly reduced when $U = V$ is enforced. This is because, in time-irreversible processes, predictive features are not necessarily predictable, and vice versa. However, because this work did not compare CCA (the optimal ${U \neq V}$ method) to DCA (the optimal ${U = V}$ method), the results are overly pessimistic. We repeated the analysis of \citep{tegmark2019optimal} using both the noisy Lorenz embedding of Fig.~\ref{fig:lorenz} as well as a system of coupled oscillators that was used in \citep{tegmark2019optimal}. For both systems, the single projection found by DCA captured almost as much past-future mutual information as the pair of projections found by CCA (Fig.~\ref{fig:methods_comparison}B,C). This suggests that while predictive and predictable features are different in general, shared past and future features might suffice for capturing most of the past-future mutual information in a certain systems. Identifying and characterizing this class of systems could have important implications for prediction-based unsupervised learning techniques \cite{tegmark2019optimal, oord2018representation}.

In addition to SFA, other time-based linear dimensionality reduction methods have been proposed. Maximum Autocorrelation Factors \citep{larsen2002decomposition} is equivalent to the version of SFA described here. Complexity Pursuit \citep{hyvarinen2001complexity} and Forecastable Components Analysis \citep{goerg2013forecastable} each minimize the entropy of a nonlinear function of the projected variables. They are similar in spirit to the frequency-domain implementation of DCA, but do not maximize past-future mutual information. Several algorithms inspired by Independent Components Analysis that incorporate time have been proposed~\citep{tong1990amuse, ziehe1998tdsep, stogbauer2004least}, but are designed to separate independent dimensions in time series rather than discover a dynamical subspace with potentially correlated dimensions. Like DCA, Predictable Feature Analysis \citep{richthofer2015predictable, weghenkel2017graph} is a linear dimensionality reduction method with a prediction-based objective. However, Predictable Feature Analysis requires explicitly specifying a prediction model, whereas DCA does not assume a particular model. Moreover, Predictable Feature Analysis requires alternating optimization updates of the prediction model and the projection matrix, whereas DCA is end-to-end differentiable. Finally, DCA is related to the Past-Future Information Bottleneck \cite{creutzig2009past} (see Appendix F).

We have been made aware of two existing methods which share the name Dynamical Component(s) Analysis~\citep{thirion2003dynamical, seifert2018dynamical, korn2019dynamical}. Thematically, they share the goal of uncovering low-dimensional dynamics from time series data. \citet{thirion2003dynamical} perform a two-stage, temporal then kernelized spatial analysis. \citet{seifert2018dynamical} and \citet{korn2019dynamical} assume the observed dynamics are formed by low-dimensional latent variables with linear and nonlinear dynamics. To fit a linear approximation of the latent variables, they derive a generalized eigenvalue problem which is sensitive to same-time and one-time step correlations, i.e., the data and the approximation of its first derivative.

An alternative to objective function-based components analysis methods are generative models, which postulate a low-dimensional latent state that has been embedded in high-dimensional observation space. Generative models featuring latent states imbued with dynamics, such as the Kalman filter, Gaussian Process Factor Analysis and LFADS, have found widespread use in neuroscience (see Appendix I for comparisons of DCA with the KF and GPFA) \citep{kalman1960new, byron2009gaussian, pandarinath2018inferring}. The power of these methods lies in the fact that rich dynamical structure can be encouraged in the latent state through careful choice of priors and model structure. However, learning and inference in generative models tend to be computationally expensive, particularly in models featuring dynamics. In the case of deep learning-based methods such as LFADS, there are often many model and optimization hyperparameters that need to be tuned. In terms of computational efficiency and simplicity, DCA occupies an attractive territory between linear methods like PCA and SFA, which are computationally efficient but extract relatively simple structure, and dynamical generative models like LFADS, which extract rich dynamical structure but are computationally demanding. As a components analysis method, DCA makes the desired properties of the learned features explicit through its objective function. Finally, the ability of DCA to yield a linear subspace in which dynamics unfold may be exploited for many analyses. For example, the loadings for DCA can be studied to examine the relationship between the high- and low-dimensional variables (Appendix J).

Lastly, while DCA does not produce an explicit description of the dynamics, this is a potentially attractive property. In particular, while dynamical generative models such as the KF provide descriptions of the dynamics, they also assume a particular form of dynamics, biasing the extracted components toward this form. By contrast, DCA is formulated in terms of spatiotemporal correlations and, as result, can extract broad forms of (stationary) dynamics, be they linear or nonlinear. For example, the Lorenz attractor of Fig. \ref{fig:lorenz} is a nonlinear dynamical system.

\section{Applications to real data}
\label{sec:results}

We used DCA to extract dynamical subspaces in four high-dimensional time series datasets: (i)~multi-neuronal spiking activity of 109 single units recorded in monkey primary motor cortex (M1) while the monkey performed a continuous grid-based reaching task~\citep{o_doherty_joseph_e_2017_583331}; (ii)~multi-neuronal spiking activity of 55 single units recorded in rat hippocampal CA1 while the rat performed a reward-chasing task~\citep{mizuseki2009multi, glaser2017machine}; (iii)~multi-city temperature data from 30 cities over several years~\citep{gene2017}; and (iv)~12 variables from an accelerometer, gyroscope, and gravity sensor recording human kinematics~\citep{malekzadeh2018protecting}. See Appendix B for details. For all results, three bins of projected data were used to predict one bin of response data. Data were split into five folds, and reported $R^2$ values are averaged across folds. 

To assess the performance of DCA, we noted that subspaces which capture dynamics should be more predictive of future states than those which capture static structure. Moreover, for the motor cortical and hippocampal datasets, subspaces which capture dynamics should be more predictive of behavioral variables (cursor kinematics and rat location, respectively) than subspaces which do not, since neural dynamics are believed to underlie or encode these variables \citep{churchland2012neural, wilson1993dynamics}. Thus, we compared the abilities of subspaces found by DCA, PCA and SFA to decode behavioral variables for the motor cortical and hippocampal datasets and to forecast future full-dimensional states for the temperature and accelerometer datasets. 

\begin{figure}[htbp!]
	\centering
	\includegraphics[width=\textwidth]{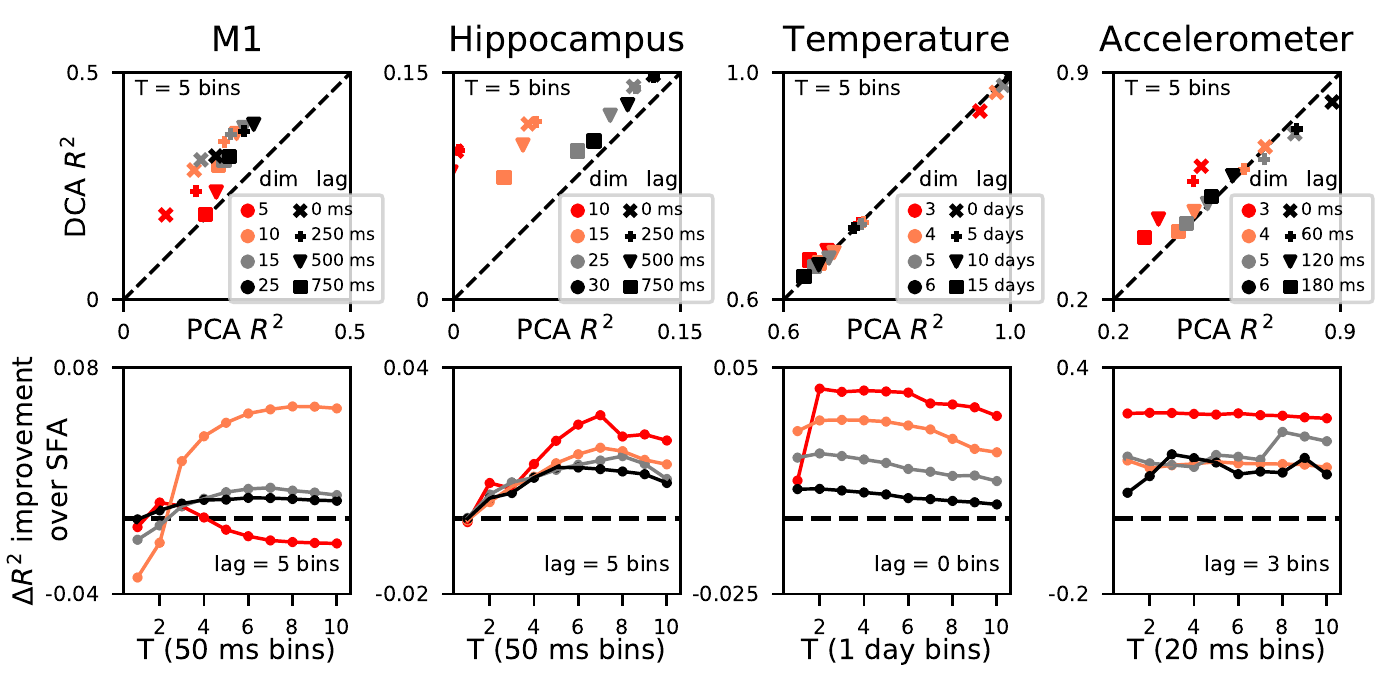}
	\caption{\textbf{DCA for prediction and forecasting.} For all panels, color indicates the projected dimensionality. For the top row, marker type indicates the lag for prediction. The top row compares held-out $R^2$ for DCA vs. PCA as a function of projection dimensionality and prediction lag. The bottom row shows the difference in held-out $R^2$ for DCA vs. SFA as a function of $T$, the past-future window size parameter for DCA. (\textbf{M1}) Predicting cursor location from projected motor cortical data. (\textbf{Hippocampus}) Predicting animal location from projected hippocampal data. (\textbf{Temperature}) Forecasting future full-dimensional temperature states from projected temperature states. (\textbf{Accelerometer}) Forecasting future full-dimensional accelerometer states from projected states.}
	\label{fig:realdata}
\end{figure}

For the motor cortical and hippocampal datasets, DCA outperformed PCA at predicting both current and future behavioral variables on held-out data (Fig.~\ref{fig:realdata}, top row). This reflects the existence of dimensions which have substantial variance, but which do not capture as much dynamical structure as other, smaller-variance dimensions. Unlike PCA, DCA is not drawn to these noisy, high-variance dimensions. In addition to demonstrating that DCA captures more dynamical structure than PCA, this analysis demonstrates the utility of DCA in a common task in neuroscience, namely, extracting low-dimensional representations of neural dynamics for visualization or further analysis (see Appendix H for forecasting results on the neural data and Appendix J for example latent trajectories and their relationship to the original measurement variables)~\citep{cunningham2014dimensionality, golub2018learning}. For the temperature dataset, DCA and PCA performed similarly, and for the accelerometer dataset, DCA outperformed PCA for the lowest-dimensional projections. The narrower performance gap between DCA and PCA on the temperature and accelerometer datasets suggests that the alignment between variance and dynamics is stronger in these datasets than in the neural data.

Assuming Gaussianity, DCA is formally superior to SFA at capturing past-future mutual information in time series which are time-irreversible and/or non-Markovian (Section \ref{section:related}). All four of our datasets possess both of these properties, suggesting that subspaces extracted by DCA might offer superior decoding and forecasting performance to those extracted by SFA. We found this to be the case across all four datasets (Fig.~\ref{fig:realdata}, bottom row). Moreover, the relative performance of DCA often became stronger as $T$ (the past-future window size of DCA) was increased, highlighting the non-Markovian nature of the data (see Appendix G for absolute $R^2$ values). This underscores the importance of leveraging spatiotemporal statistics across long timescales when extracting non-Markovian dynamical structure from data.

\section{Discussion} 
DCA retains the geometric interpretability of linear dimensionality reduction methods while implementing an information-theoretic objective function that robustly extracts dynamical structure while minimizing noise. Indeed, the subspace found by DCA may be thought of as the result of a competition between aligning the subspace with dynamics and making the subspace orthogonal to noise, as in Fig. \ref{fig:lorenz}A. Applied to neural, weather and accelerometer datasets, DCA often outperforms PCA, indicating that noise variance often dominates or is comparable to dynamics variance in these datasets. Moreover, DCA often outperforms SFA, particularly when DCA integrates spatiotemporal statistics over long timescales, highlighting the non-Markovian statistical dependencies present in these datasets. Overall, our results show that DCA is well-suited for finding dynamical subspaces in time series with structural attributes characteristic of real-world data.

Many extensions of DCA are possible. Since real-world data generation processes are generally non-stationary, extending DCA for non-stationary data is a key direction for future work. For example, non-stationary data may be segmented into windows such that the data are approximately stationary within each window~\cite{costa2019adaptive}. In general, the subspace found by DCA includes contributions from all of the original variables. For increased interpretability, DCA could be optimized with an $\ell^1$ penalty on the projection matrix $V$~\cite{tibshirani1996regression} to identify a small set of relevant features, e.g., individual neurons or stocks \cite{mahoney2009cur}. Both the time- and frequency-domain implementations of DCA may be made differentiable in the input data, opening the door to extensions of DCA that learn nonlinear transformations of the input data, including kernel-like dimensionality expansion, or that use a nonlinear mapping from the high- to low-dimensional space, including deep architectures. Since DCA finds a linear projection, it can also be kernelized using the kernel trick. The DCA objective could also be used in recurrent neural networks to encourage rich dynamics. Finally, dimensionality reduction via DCA could serve as a preprocessing step for time series analysis methods which scale unfavorably in the dimensionality of the input data, allowing such techniques to be scaled to high-dimensional data.

\section*{Acknowledgements}
D.G.C. and K.E.B. were funded by LBNL Laboratory Directed Research and Development. We thank the Neural Systems and Data Science Lab and Laurenz Wiskott for helpful discussion.

\vspace{\fill}
\pagebreak

\vspace{1em} 
\textbf{\huge{Appendix}}
\appendix

\section{Implementation, optimization \& time complexity}
DCA was implemented in PyTorch and optimized using the L-BFGS-B algorithm in SciPy~\citep{paszke2017pytorch, zhu1997algorithm, jones2001scipy}. All analysis was run on a desktop computer. To make optimization more stable, we included a penalty term in the loss function which encourages the projection matrix $V$ to have orthonormal columns during optimization. The final loss is given by
\begin{equation}
    \mathcal{L}_\text{DCA} = -I^\text{pred}_T\left(Y\right) + \lambda\left\lVert V^TV - I \right\rVert_F^2
\end{equation}
where $\lambda > 0$. The orthonormality penalty is always exactly zero at the end of optimization since mutual information is invariant under invertible transformations of either of its arguments, so we can always transform $V$ to have orthonormal columns without changing the predictive information. Because the loss function is non-convex, we typically perform 5 random initializations.

For large spatial or temporal dimensions with large correlations, the spatiotemporal covariance matrix was occasionally not positive-definite. In these cases, we added a constant to the diagonal of $C_0$, the same-time covariance matrix, so that the smallest eigenvalue was 1e-6 (equivalent to adding a small amount of uncorrelated noise to all dimensions for all times).

To fit DCA for a dataset with total length $T_{\text{tot}}$, dimensionality $n$ and projection dimensionality $d$ using past and future windows of length $T$, the first step is to compute the spatiotemporal covariance matrix, which is $\mathcal{O}\left(n^2T^2 T_{\text{tot}} \right)$. The time complexity of DCA's optimization procedure does not scale in the total amount of data since the objective references the data only through this spatiotemporal covariance matrix which is computed prior to optimization. Each evaluation of the objective, or its gradient, requires computing $2T$ quadratic products of the form $V^T C_{\Delta t} V$, each of which is $\mathcal{O}\left(n^2d + nd^2 \right)$, as well as the log-determinants, which are $\mathcal{O}\left(T^3 d^3 \right)$. Altogether, each evaluation is $\mathcal{O}\left(Tn^2 d + T n d^2 + T^3d^3\right)$.  For all datasets we considered, optimization time dominated the time required to compute the spatiotemporal covariance matrix.

\section{Datasets and preprocessing}

\subsection{Lorenz attractor synthetic data}
The governing equations of the Lorenz attractor are~\citep{strogatz2018nonlinear}
\begin{equation}
\begin{split}
    \dot{x} &= \sigma\left(y - x\right) \\ 
    \dot{y} &= x\left(\rho - z\right) - y \\ 
    \dot{z} &= xy - \beta z.
\end{split}
\end{equation}
In all appearances of the Lorenz attractor in the main text, we used the parameters ${\sigma=10}$, ${\beta=\frac{8}{3}}$ and ${\rho = 28}$, which place the system in the chaotic regime. We used an integration time step of ${\Delta t=5 \times 10^{-3}}$, then downsampled the data by a factor of 5.

For the {30-dimensional} noisy embedded Lorenz attractor used in Figures 1-3, we first generated three-dimensional Lorenz data as described above, then embedded the dynamics into 30-dimensional space via a random orthogonal embedding. We then added Gaussian white noise, which we parameterized by the eigenspectrum and eigenvectors of its covariance matrix. In particular, the eigenspectrum was given by
\begin{equation}
    \lambda_j = \sigma^2 \exp\left( - \frac{2 j}{d_\text{noise}} \right)
\end{equation}
where $\sigma^2$ controls the overall amount of noise and $d_\text{noise}$ controls the effective dimensionality of the noise. In all cases, we used $d_\text{noise} = 7$. In Fig. 1, the parameter $\sigma^2$, which is also the variance of the first principal component of of the noise, was varied to obtain different SNR values. In Figures 2 and 3, $\sigma^2$ was fixed to achieve an SNR of unity. The eigenvectors of the noise covariance were chosen uniformly at random with the constraint the leading $d_\text{noise}$ eigenvectors had close-to-median principal angles with respect to the subspace containing the Lorenz attractor.

\subsection{Monkey motor cortical dataset}
\citet{o_doherty_joseph_e_2017_583331} released multi-electrode spiking data for both M1 and S1 for two monkeys during a continuous grid-based reaching task. We used M1 data from the subject ``Indy'' (specifically, we used the file ``indy\_20160627\_01.mat''). We discarded single units with fewer than 5,000 spikes, leaving 109 units. We binned the spikes into non-overlapping bins (100 ms in Fig. 3, 50 ms in Figures 2 and 4), square-root transformed the data and mean-centered the data using a sliding window 30 s in width. 

\subsection{Rat hippocampal data}
\citet{glaser2017machine} released the original data of~\citet{mizuseki2009multi} (dataset ``hc2'', session ``ec014.333''). The data consist of 93 minutes of extracellular recordings from layer CA1 of dorsal hippocampus while a rat chased rewards on a square platform. We discarded single units with fewer than 10 spikes, leaving 55 units. We binned the spikes into non-overlapping 50 ms bins, then square-root transformed the data.

\subsection{Temperature dataset}
The temperature dataset consists of hourly temperature data for 30 U.S. cities over a period of 7 years from \url{OpenWeatherMap.org}~\citep{gene2017}. We downsampled the data by a factor of 24 to obtain daily tempeartures. 

\subsection{Accelerometer dataset}
\citet{malekzadeh2018protecting} released accelerometer data which records roll, pitch, yaw, gravity $\{x, y, z\}$, rotation $\{x, y, z\}$ and acceleration $\{x, y, z\}$ for a total of 12 kinematic variables. The sampling rate is 50 Hz. We used the file ``sub\_19.csv'' from ``A\_DeviceMotion\_data.zip''.

\section{Frequency-domain predictive information computation}
Consider a one-dimensional discrete-time Gaussian process $Y$ with zero mean and autocovariance function $f(\Delta t)$. We can use Theorem 2.5 from \citet{li1996model} (see also \citep{li2006some}) to compute the predictive information of $Y$ in terms of the discrete-time Fourier transform (DTFT) of $f(\Delta t)$. Specifically, this theorem says that when the asymptotic predictive information $I^\text{pred}_{T \rightarrow \infty}(Y)$ is finite, we have
\begin{equation}
	I^\text{pred}_{T \rightarrow \infty}(Y) = \sum_{k=1}^{\infty} k b_k^2,
\end{equation}
as stated in the main text. The numbers $\{b_k\}$ are called the cepstrum coefficients of $Y$. They comprise the DTFT of the logarithm of the DTFT of $f(\Delta t)$:
\begin{equation}
    b_{k} = \frac{1}{2\pi} \int_{-\pi}^\pi d\lambda e^{-i \lambda k} \log \tilde{f}(\lambda), \:\:\: \tilde{f}(\lambda) = \sum_{k=-\infty}^\infty e^{-i \lambda k} f(k).
\end{equation}

In practice, rather than directly computing the autocovariance function $f(\Delta t)$ of $Y$ and taking its DTFT, we compute the power spectral density of $Y$, which is equivalent to the DTFT of $f(\Delta t)$. Specifically, we use the FFT in conjunction with a window function to compute the power spectral density in many length-$2T$ windows of $Y$, then average the results together. If $f(\Delta t)$ falls off to zero with a timescale $\tau \ll 2T$, then this method computes the full asymptotic predictive information $I^\text{pred}_{T \rightarrow \infty}(Y)$ in the limit of infinite samples. If $f(\Delta t)$ does not fall off this quickly, then the window function effectively forces the autocovariance to decay to zero at $\Delta t = \pm 2T$, yielding a regularized estimate of $I^\text{pred}_T(Y)$.

\section{Asymtotic predictive information derivations}
\subsection{Exponential autocovariance}
Let $Y_1$ be a discrete-time Gaussian process whose autocovariance function $f_1(\Delta t)$ is an exponential:
\begin{equation}
f_1(\Delta t)= \exp\left(- \left| \frac{\Delta t }{ \tau} \right| \right).
\end{equation}
It is easy to show that $f_1(\Delta t)$ is the autocovariance function of an AR(1) process
\begin{equation}
y_t = A y_{t-1} + e_t
\end{equation}
where $\left\langle e_t^2 \right\rangle_t = \Omega^2$ and $ \left\langle e_{t}e_{t+{\Delta t}}\right\rangle_t = 0$ for $|{\Delta t}| \geq 1$. The autocovariance of this process is
\begin{equation}
\mathbf{E}[y_t y_{t + {\Delta t}}] = \frac{\Omega^2}{1 - A^2} A^{-|{\Delta t}|},
\end{equation}
and if we set $\Omega^2 = 1 - e^{\frac{2}{\tau}}$ and $A = e^{\frac{1}{\tau}}$ we have
\begin{equation}
\mathbf{E}[y_t y_{t + {\Delta t}}] = e^{-\frac{|k|}{\tau}} = f_1({\Delta t}).
\end{equation}
Since this process is Markovian, $I^\text{pred}_{T}\left(Y_1\right)$ is simply the mutual information between two consecutive time steps for all $T \geq 1$. Thus, we have
\begin{equation}
I^\text{pred}_{T \rightarrow \infty}\left(Y_1\right) = -\frac{1}{2}\log\left(1 - f_1(1)^2 \right) = 
-\frac{1}{2}\log\left(1 - e^{-\frac{2}{\tau}} \right)
\end{equation}
and for $\tau \gg 1$, this becomes
\begin{equation}
I^\text{pred}_{T \rightarrow \infty}\left(Y_1\right) = \frac{1}{2}\log \frac{\tau}{2}.
\end{equation}

\subsection{Squared-exponential autocovariance}
Let $Y_2$ be a discrete-time Gaussian process whose autocovariance function $f_2(\Delta t)$ is a squared-exponential:
\begin{equation}
f_2(\Delta t)= \exp\left(- \frac{\Delta t^2 }{ \tau^2} \right).
\end{equation}
We will compute the predictive information using the cepstrum coefficients. The DTFT $\tilde{f}_2(\lambda)$ of $f_2(\Delta t)$ is
\begin{equation}
\tilde{f}_2(\lambda) = \sum_{k=-\infty}^\infty \cos(k \lambda) f_2(k) = 
\sqrt{\pi} \tau e^{-\frac{1}{4}\lambda^2 \tau^2} \sum_{k=-\infty}^\infty e^{-\tau^2 \left(k^2 \pi^2 + k \pi \lambda \right) }.
\end{equation}
Taking the log gives
\begin{equation}
\log \tilde{f}_2(\lambda) = \frac{1}{2}\sqrt{\pi} + \log \tau - \frac{1}{4}\lambda^2 \tau^2 
+ \log\left( 1 + e^{-\tau^2(\pi^2 + \pi \lambda)} + e^{-\tau^2(\pi^2 - \pi \lambda)} + \cdots \right).
\end{equation}
For $\tau \gg 1$, the logarithmic term on the RHS is approximately $\log 1 = 0$ (this is not true close to the endpoints $\lambda = \pm \pi$ where this term is approximately $\log 2$, however $\log \tilde{f}_2(\lambda)$ also contains a term quadratic in $\tau$ which dominates the logarithmic term close to $\lambda = \pm \pi$). Taking another DTFT gives the cepstrum coefficients:
\begin{equation}
b_k \approx \frac{1}{\pi}\int_{0}^\pi\cos(k \lambda) \left( \frac{1}{2}\sqrt{\pi} + \log \tau - \frac{1}{4}\lambda^2 \tau^2  \right)  d\lambda 
= -\frac{\tau^2 \cos(k \pi)}{2 k^2}.
\end{equation}
Thus, we have
\begin{equation}
I^\text{pred}_{T \rightarrow \infty}(Y_2) = \frac{1}{2}\sum_{k=1}^\infty k b_k^2
= \frac{\tau^4}{8} \sum_{k=1}^\infty \frac{1}{k^3} = \frac{\zeta(3)}{8} \tau^4 \approx 0.15026 \times \tau^4
\end{equation}
where $\zeta$ is the Riemann zeta function.

\section{Note on Slow Feature Analysis}
If we allow for one-dimensional projections of $X$ with negative one-time step autocorrelation $\rho_1 < 0$, then the one time-step mutual information $I_1$ is non-monotonically related to $\rho_1$ across different projections according to ${I_1 = - \frac{1}{2} \log\left(1 - \rho_1^2 \right)}$. As a result, SFA is no longer guaranteed to coincide with DCA for $d = 1$, nor for time-reversible processes with $d \geq 1$. However, if we modify SFA to order the projections according to decreasing $\rho_1^2$, rather than decreasing $\rho_1$, then these guarantees are restored. The notion that slowness (${\rho_1 > 0}$) and fastness (${\rho_1 < 0}$) are equivalent in the eyes of mutual information was pointed out by \citet{creutzig2008predictive}, who presented the information-theoretic interpretation of SFA for time-reversible processes.



\section{Note on the Past-Future Information Bottleneck}

The problem of finding representations of time series that optimally capture past-future mutual information has been studied in the context of the Information Bottleneck (IB), a method for compressing data in a way that retains its relevant aspects \cite{tishby2000information}. In particular, the Past-Future Information Bottleneck (PFIP) seeks a compressed representation $Y$ of $X_\text{past}$ that has maximal mutual information with $X_\text{future}$, subject to a fixed amount of compression \cite{creutzig2008predictive, creutzig2009past, amir2015past}. This corresponds to a variational problem with the Lagrangian
\begin{equation}
    \mathcal{L}_\text{PFIB} = I\left(X_\text{past} ; Y \right) - \beta I\left(Y ; X_\text{future} \right),
\end{equation}
where $\beta$ controls the tradeoff between compression of $X_\text{past}$ and prediction of $X_\text{future}$. The most fundamental difference between the PFIB and DCA is that the PFIB compresses only the past, whereas DCA compresses both the past and the future by projecting to a lower-dimensional space. However, there is nonetheless a case in which the PFIB and DCA solutions coincide.

When the ``observed'' and ``relevant'' variables in an IB problem ($X_\text{past}$ and $X_\text{future}$ in the PFIB) are jointly Gaussian, then the solution is closely related to CCA \cite{chechik2005information}. As $\beta$ is increased, the solution undergoes a cascade of structural phase transitions which increase the dimensionality of the compressed representation, i.e., the number of CCA components retained. For one-time step past and future windows, the PFIB solution is to retain the top-$d$ left singular vectors of $C_0^{-1/2}C_1C_0^{-1/2}$, where $d$ is determined by $\beta$. For time-reversible processes, the PFIB solution coincides with that of DCA (and with SFA \cite{creutzig2008predictive}). This is potentially surprising, since the PFIB maximizes $I\left(y_t ; x_{t+1}\right)$ while DCA maximizes $I\left(y_t ; y_{t+1}\right)$. Put differently, in Gaussian processes with time-reversal symmetry, the features of the past which are the most self-predictive are also the most predictive overall. For time-irreversible processes, DCA, SFA and the PFIB admit mutually distinct solutions: we show in the main text that DCA and SFA disagree, and the PFIB solution must be different from those of both DCA and SFA since the solutions to both of these methods are invariant under time-reversal transformations while the solution to the PFIB is not. 

\section{Absolute \texorpdfstring{$\mathbf{R^2}$}{R2} values}

\begin{figure}[htbp!]
	\centering
	\includegraphics[width=5.2in]{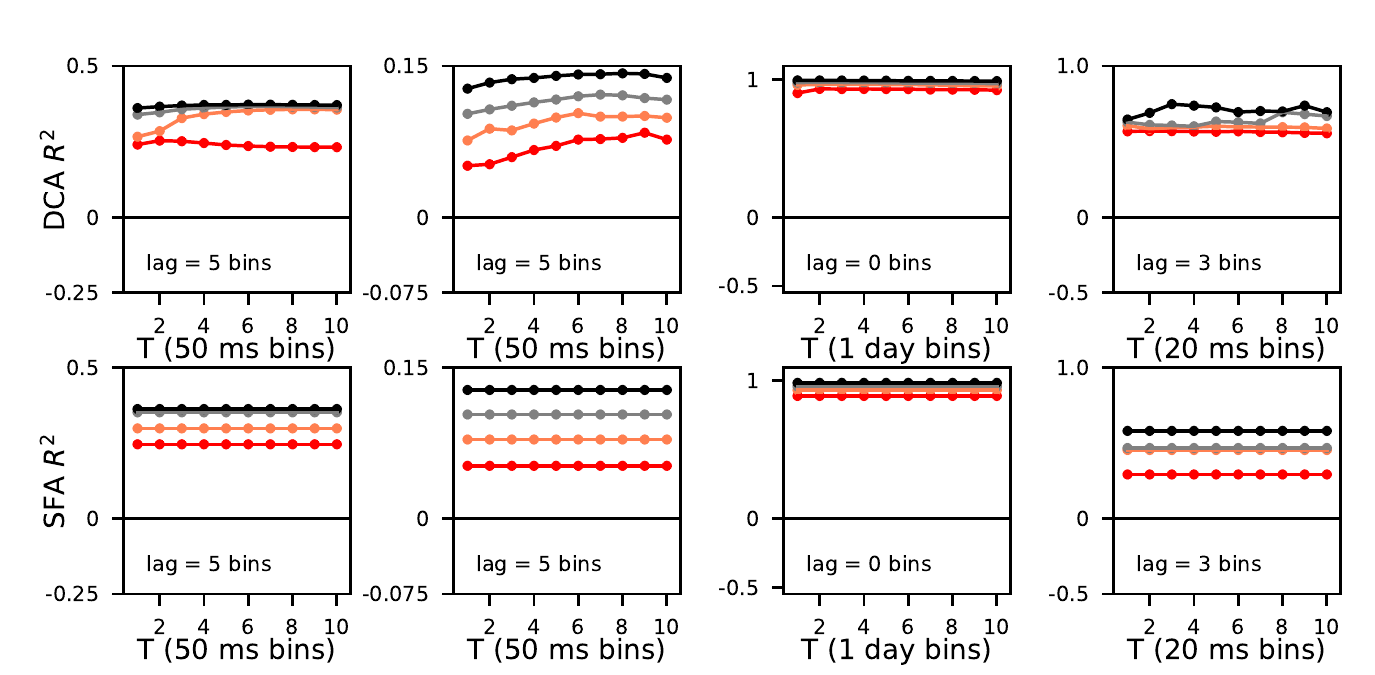}
	\caption{\textbf{Absolute $\mathbf{R^2}$.} Absolute held-out $R^2$ values for the DCA vs. SFA comparisons in Fig~4 of the main text. See main text for legends.}
	\label{fig:absolute_r2}
\end{figure}

Fig~\ref{fig:absolute_r2} shows the absolute held-out $R^2$ values for the DCA vs. SFA comparison in Fig~4 of the main text. Note that SFA does not depend on $T$.

\section{Neural Forecasting}
\begin{figure}[htbp!]
	\centering
	\includegraphics[width=3.25in]{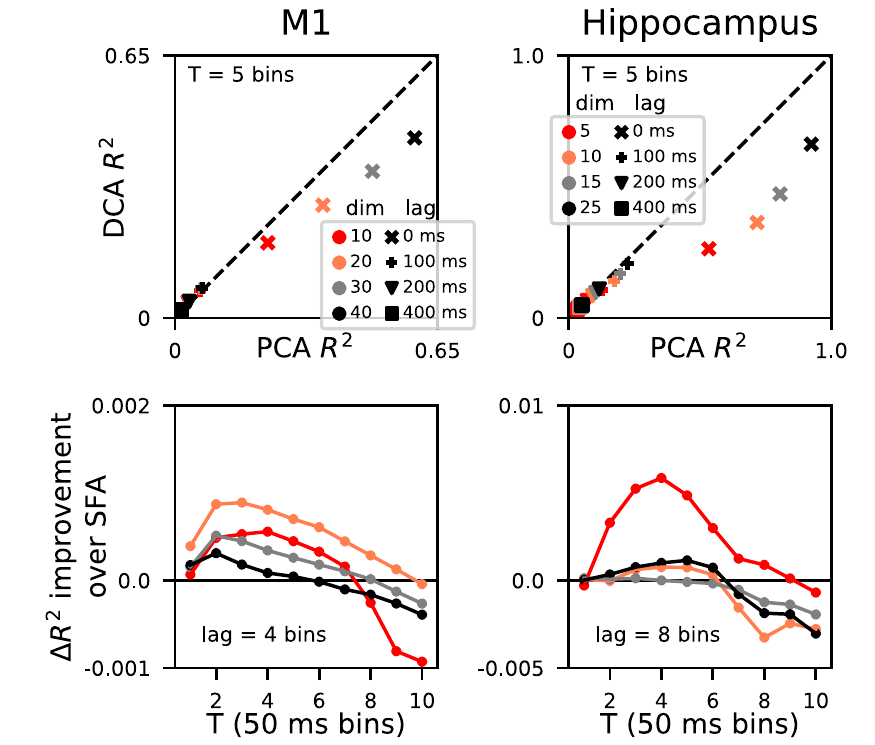}
	\caption{\textbf{Neural forecasting.} For the M1 and hippocampus datasets, the held-out $R^2$ is shown across dimension and lags.}
	\label{fig:neural_forecast}
\end{figure}

For the M1 and hippocampus datasets in Fig~4 of the main text, we also ran the forecasting analysis in which the projected neural state is used to predict future full-dimensional neural states. Fig~\ref{fig:neural_forecast} shows comparisons of DCA with PCA and SFA. At short time lags, PCA is expected to have higher $R^2$ due to the optimality of PCA at capturing the Frobenius norm. At longer time lags, DCA, PCA, and SFA all have relatively low predictive power.

\section{Kalman Filter and Gaussian Process Factor Analysis}
Two popular dynamical generative models used to infer latent dynamics from time series data are the Kalman Filter (KF) and Gaussian Process Factor Analysis (GPFA). The KF assumes that a latent time series with linear, Gaussian dynamics has been linearly embedded into observation space with Gaussian observation noise. Similarly, GPFA assumes that a latent time series whose components are independent Gaussian processes with a common kernel, but independent timescales, has been linearly embedded into observation space with Gaussian observation noise. In both cases, the model parameters are fit using the expectation-maximization (EM) algorithm \cite{ghahramani1996parameter, byron2009gaussian}. To infer the latent state at time $t$, each model admits both a causal procedure, which uses observations at times ${t' \in [1, \ldots, t]}$, as well as a non-causal procedure, which uses observations at times ${t' \in [1, \ldots, T]}$. For the KF, the causal and non-causal procedures are called Kalman filtering and smoothing, respectively. For each of the four datasets analyzed in the main text, we inferred latent states using the KF and GPFA using both causal and non-causal inference procedures using 5-fold cross validation. Note that while both the causal and non-causal procedures for each method use many observations to infer each time step of the latent dynamics, DCA uses only one. Thus, performance comparisons between DCA and the KF or GPFA are somewhat ill-posed since the latent factors for DCA incorporate less information. However, between the causal and non-causal inference procedures, the causal procedures are better suited for comparison to DCA since their resulting latent factors do not incorporate future observations.

\begin{figure}[htbp!]
	\centering
	\includegraphics[width=\textwidth]{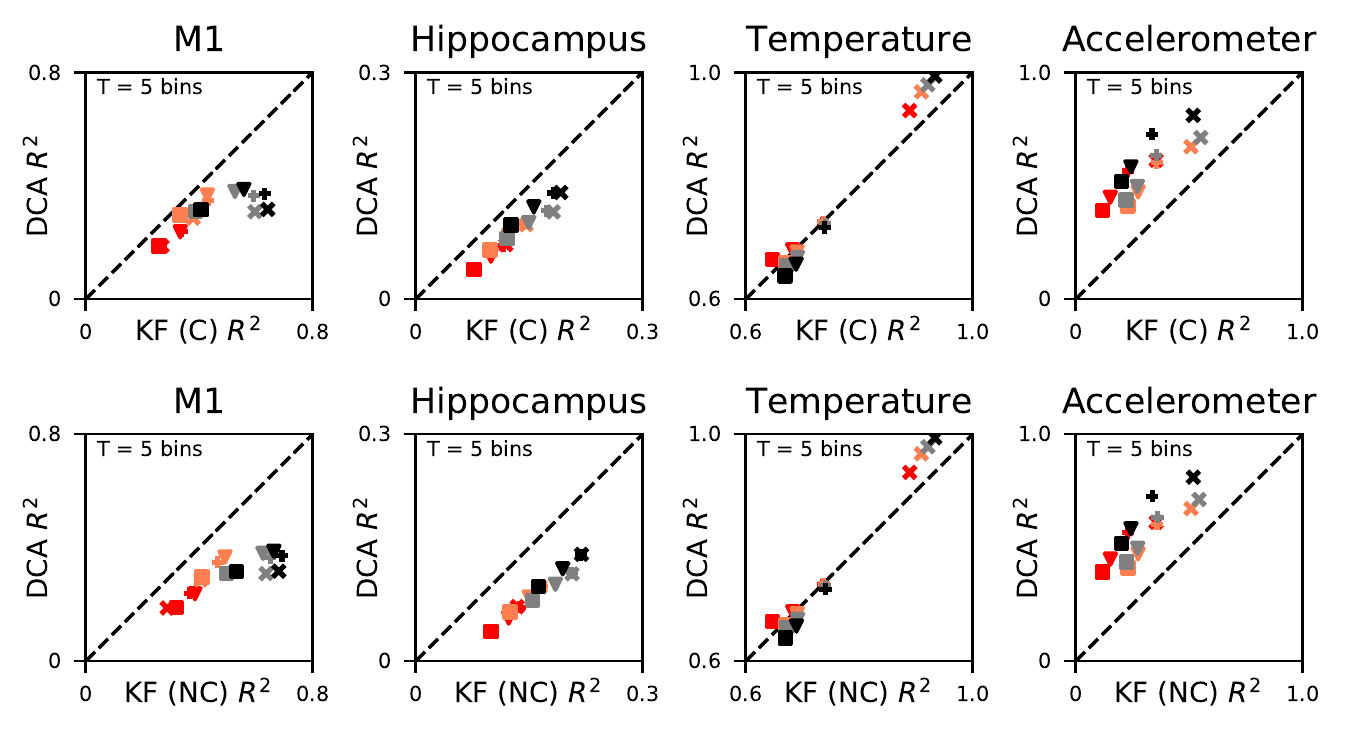}
	\caption{\textbf{DCA vs. the Kalman filter.} For all panels, color indicates the number of factors and marker type indicates the lag for prediction (see Fig. 4 of the main text for legends). Each panel compares held-out $R^2$ for DCA vs. PCA as a function of the number of factors and prediction lag. Top row: causal inference procedure (Kalman filtering). Bottom row: non-causal inference procedure (Kalman smoothing).}
	\label{fig:kf}
\end{figure}

The KF model was fit using its EM algorithm, as derived in \citet{ghahramani1996parameter}. During the E step, which computes the parameters of the Gaussian distribution over the latent states using forward and backward passes, we employed a steady-state optimization in which we did not recompute matrices which had converged to their steady-state values during each pass\footnote{Our Kalman filter EM code is available at: \url{https://github.com/davidclark1/FastKF}}. Latent factors inferred using both the causal and non-causal inference procedures for the KF performed better than factors extracted using DCA at decoding behavioral variables from neural data, and the performance gap was largest for large numbers of factors and short time lags (Fig. \ref{fig:kf}, M1 and Hippocampus). For the non-neural datasets, factors extracted using DCA generally had better forecasting performance, particularly for the accelerometer dataset (Fig. \ref{fig:kf}, Temperature and Accelerometer).

\begin{figure}[htbp!]
	\centering
	\includegraphics[width=\textwidth]{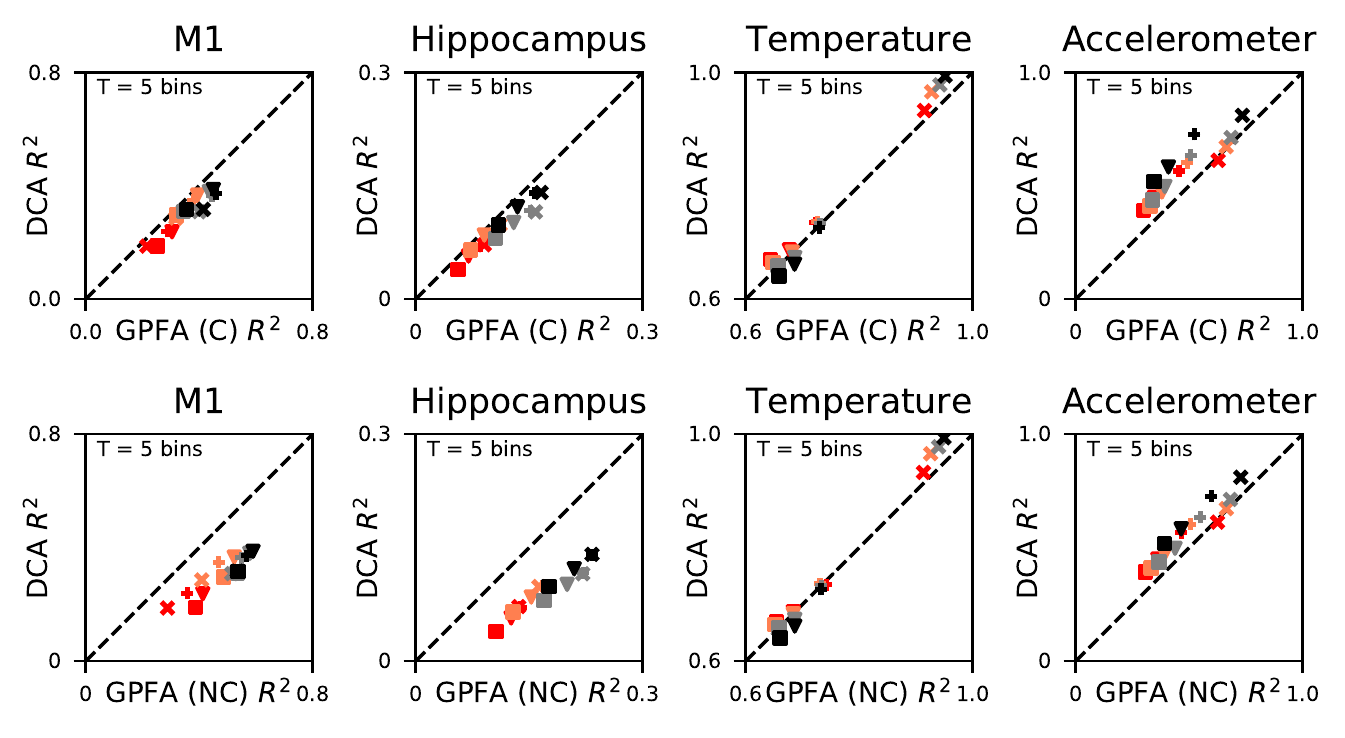}
	\caption{\textbf{DCA vs. GPFA.} For all panels, color indicates the number of factors and marker type indicates the lag for prediction (see Fig. 4 of the main text for legends). Each panel compares held-out $R^2$ for DCA vs. PCA as a function of the number of factors and prediction lag. Top row: causal inference procedure. Bottom row: non-causal inference procedure.}
	\label{fig:gpfa}
\end{figure}

We used the MATLAB code accompanying \citet{byron2009gaussian} for EM and inference in the GPFA model. For all datasets, the data were segmented into non-overlapping ``trials'' of 100 time steps, which provided substantial speedups. For performance evaluation, factors extracted by DCA were segmented to have the same trial structure as those inferred using GPFA and care was taken to not evaluate decoding or forecasting performance across trial boundaries. EM was initialized using factor analysis and run for 300 iterations, which we confirmed was adequate for convergence based on inspection of the log-likelihood over training. The provided implementation of GPFA automatically segmented our trials of length 100 into shorter segments of length 20 during fitting. Latent factors inferred using the causal procedure for GPFA performed slightly better than factors extracted using DCA at decoding behavioral variables from neural data (Fig. \ref{fig:gpfa}, M1 and Hippocampus, top row) while the performance gap for the non-causal inference procedure was larger (Fig. \ref{fig:gpfa}, M1 and Hippocampus, bottom row). For the non-neural datasets, factors extracted using DCA performed slightly better than those inferred using either the causal or non-causal procedures for GPFA (Fig. \ref{fig:gpfa}, Temperature and Accelerometer).

\section{Inferred states and leverage scores}

\begin{figure}[htbp!]
	\centering
	\includegraphics[width=4in]{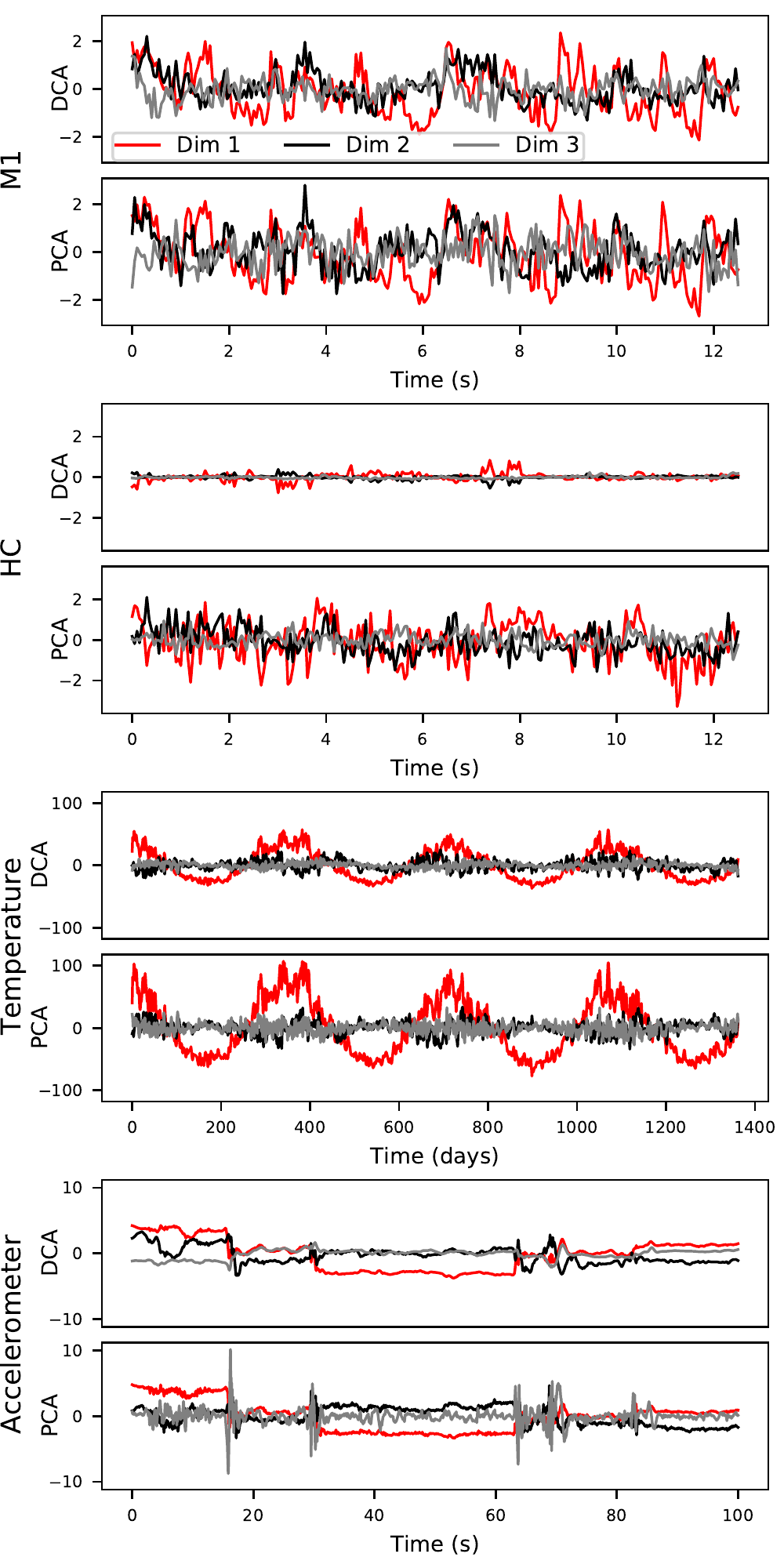}
	\caption{\textbf{Inferred latent states for DCA vs. PCA.} For each dataset, the projections into a three-dimensional space are shown for DCA and PCA. The projections have been ordered by variance explained in the subspace.}
	\label{fig:inferred_states}
\end{figure}

For each of the four datasets considered, we visualized the extracted latent states after three-dimensional projections found using DCA and PCA. Fig~\ref{fig:inferred_states} shows the projections from DCA (top panels in pairs) compared to the projections from PCA (bottom panel in pairs). Since DCA yields a subspace rather than an ordered sequence of components, we transformed the DCA projections using PCA so that the DCA components were ordered by variance explained, making the comparison to PCA more clear. The Spearman's rank-correlation between the top 3 components of DCA and PCA are: M1 (0.98, 0.76, 0.3), HC (0.42, 0.15, 0.12), Temperature (1.0, 0.93, 0.78) and {Accelerometer (0.89, 0.88, -0.11)}. As expected, the DCA projections are typically lower amplitude compared to the PCA projections. This is exaggerated in the hippocampus dataset. In the accelerometer dataset, the DCA projections are smoother than their PCA counterparts.

The DCA and PCA subspaces can also be compared through their leverage scores~\cite{mahoney2009cur} which measure the level of alignment between a subspace and the original measurement axes. Given an orthonormal basis $V \in \mathbb{R}^{n \times d}$ for a $d$-dimensional subspace, the leverage score for measurement axis $j$ is
\begin{equation}
    \pi_j = \frac{1}{d}\sum_{i=1}^d\left(v_{ij}\right)^2
\end{equation}
where, from the definition, ${\sum_j\pi_j=1}$ and ${0\le \pi_j \le 1}$. This means that $\pi_j$ can be thought of as a distribution.

\begin{figure}[htbp!]
	\centering
	\includegraphics[width=4.75in]{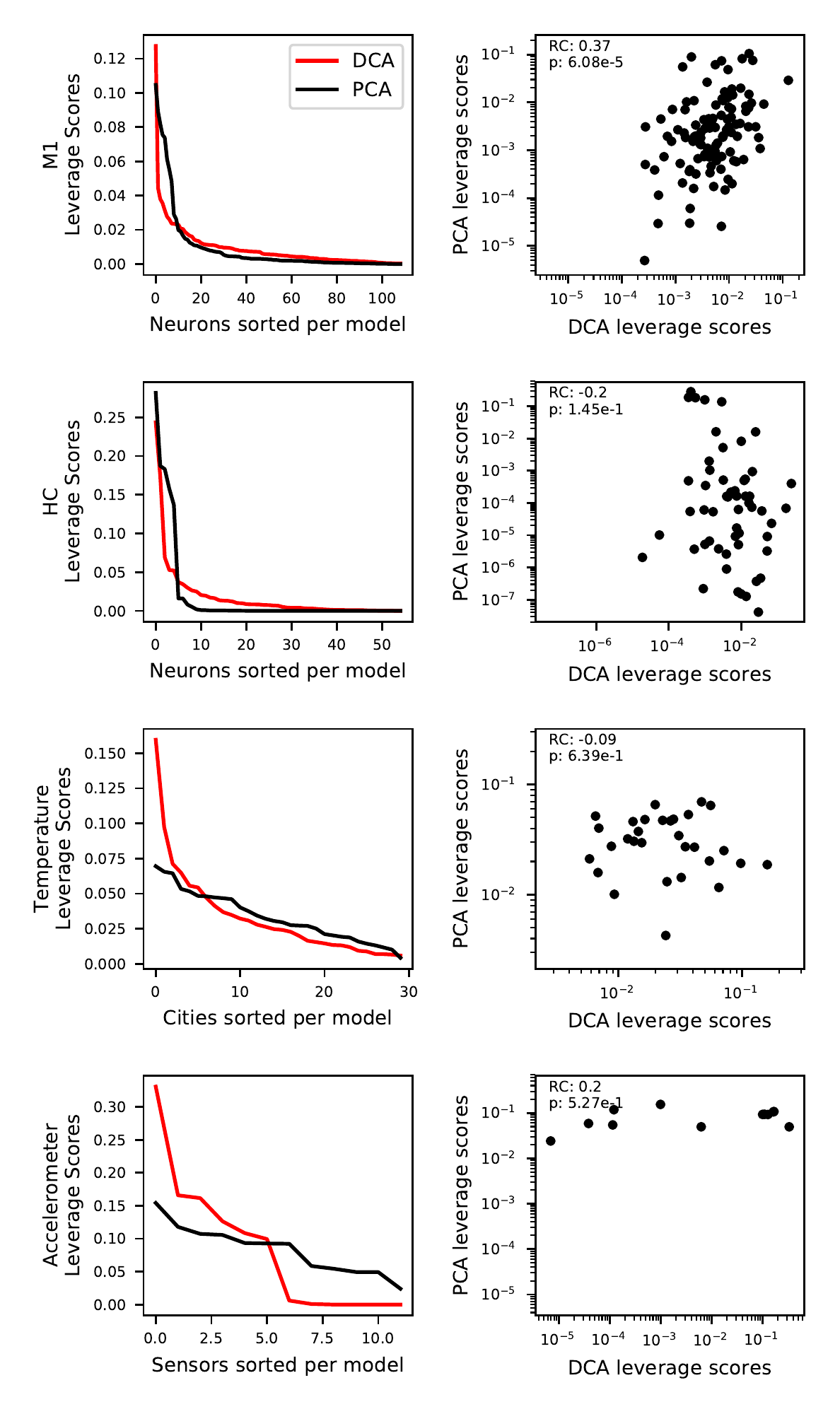}
	\caption{\textbf{Leverage scores for the DCA and PCA subspaces.} The left panels show the distribution of leverage scores when they are sorted individually for DCA and PCA. The right panels show the DCA leverage scores vs. the PCA leverage scores per measurement axis (log-log scale). The Spearman's rank-correlation (RC) and significance is inset for each dataset.}
	\label{fig:leverage}
\end{figure}

Fig~\ref{fig:leverage} shows the sorted leverage scores and a comparison of the DCA and PCA leverage scores across measurement axes. For all datasets except for the hippocampus dataset, the DCA leverage scores have a sharper peak compared to the PCA leverage scores. For the neural datasets, the DCA leverage scores have a larger tail compared to the PCA leverage scores. On the temperature and accelerometer datasets, DCA has a few measurement axes with large leverage scores and a slowly decaying tail while the PCA leverage scores decay approximately linearly. Only on the M1 dataset are the DCA and PCA leverage scores significantly correlated as measured by the Spearman's rank-correlation (RC).

\vspace{\fill}

\pagebreak 
\bibliographystyle{unsrtnat}
\bibliography{refs}

\begin{thebibliography}{62}
\providecommand{\natexlab}[1]{#1}
\providecommand{\url}[1]{\texttt{#1}}
\expandafter\ifx\csname urlstyle\endcsname\relax
  \providecommand{\doi}[1]{doi: #1}\else
  \providecommand{\doi}{doi: \begingroup \urlstyle{rm}\Url}\fi

\bibitem[Gao and Ganguli(2015)]{gao2015simplicity}
Peiran Gao and Surya Ganguli.
\newblock On simplicity and complexity in the brave new world of large-scale
  neuroscience.
\newblock \emph{Current opinion in neurobiology}, 32:\penalty0 148--155, 2015.

\bibitem[Cunningham and Ghahramani(2015)]{cunningham2015linear}
John~P Cunningham and Zoubin Ghahramani.
\newblock Linear dimensionality reduction: {Survey}, insights, and
  generalizations.
\newblock \emph{The Journal of Machine Learning Research}, 16\penalty0
  (1):\penalty0 2859--2900, 2015.

\bibitem[Mahoney and Drineas(2009)]{mahoney2009cur}
Michael~W Mahoney and Petros Drineas.
\newblock {CUR} matrix decompositions for improved data analysis.
\newblock \emph{Proceedings of the National Academy of Sciences}, 106\penalty0
  (3):\penalty0 697--702, 2009.

\bibitem[Mikolov et~al.(2013)Mikolov, Chen, Corrado, and
  Dean]{mikolov2013efficient}
Tomas Mikolov, Kai Chen, Greg Corrado, and Jeffrey Dean.
\newblock Efficient estimation of word representations in vector space.
\newblock \emph{arXiv preprint arXiv:1301.3781}, 2013.

\bibitem[Doersch et~al.(2015)Doersch, Gupta, and
  Efros]{doersch2015unsupervised}
Carl Doersch, Abhinav Gupta, and Alexei~A Efros.
\newblock Unsupervised visual representation learning by context prediction.
\newblock In \emph{Proceedings of the IEEE International Conference on Computer
  Vision}, pages 1422--1430, 2015.

\bibitem[Kim et~al.(2016)Kim, Jernite, Sontag, and Rush]{kim2016character}
Yoon Kim, Yacine Jernite, David Sontag, and Alexander~M Rush.
\newblock Character-aware neural language models.
\newblock In \emph{Thirtieth AAAI Conference on Artificial Intelligence}, 2016.

\bibitem[Marzen and Crutchfield(2017)]{marzen2017nearly}
Sarah~E Marzen and James~P Crutchfield.
\newblock Nearly maximally predictive features and their dimensions.
\newblock \emph{Physical Review E}, 95\penalty0 (5):\penalty0 051301, 2017.

\bibitem[McAllester(2018)]{mcallester2018information}
David McAllester.
\newblock {Information Theoretic Co-Training}.
\newblock \emph{arXiv preprint arXiv:1802.07572}, 2018.

\bibitem[Oord et~al.(2018)Oord, Li, and Vinyals]{oord2018representation}
Aaron van~den Oord, Yazhe Li, and Oriol Vinyals.
\newblock Representation learning with {Contrastive Predictive Coding}.
\newblock \emph{arXiv preprint arXiv:1807.03748}, 2018.

\bibitem[Bouchard and Brainard(2016)]{bouchard2016auditory}
Kristofer~E Bouchard and Michael~S Brainard.
\newblock Auditory-induced neural dynamics in sensory-motor circuitry predict
  learned temporal and sequential statistics of birdsong.
\newblock \emph{Proceedings of the National Academy of Sciences}, 113\penalty0
  (34):\penalty0 9641--9646, 2016.

\bibitem[Tishby et~al.(2000)Tishby, Pereira, and Bialek]{tishby2000information}
Naftali Tishby, Fernando~C Pereira, and William Bialek.
\newblock The information bottleneck method.
\newblock \emph{arXiv preprint physics/0004057}, 2000.

\bibitem[Bialek et~al.(2001)Bialek, Nemenman, and
  Tishby]{bialek2001predictability}
William Bialek, Ilya Nemenman, and Naftali Tishby.
\newblock Predictability, complexity, and learning.
\newblock \emph{Neural computation}, 13\penalty0 (11):\penalty0 2409--2463,
  2001.

\bibitem[Palmer et~al.(2015)Palmer, Marre, Berry, and
  Bialek]{palmer2015predictive}
Stephanie~E Palmer, Olivier Marre, Michael~J Berry, and William Bialek.
\newblock Predictive information in a sensory population.
\newblock \emph{Proceedings of the National Academy of Sciences}, 112\penalty0
  (22):\penalty0 6908--6913, 2015.

\bibitem[Pearson(1901)]{pearson1901liii}
Karl Pearson.
\newblock On lines and planes of closest fit to systems of points in space.
\newblock \emph{The London, Edinburgh, and Dublin Philosophical Magazine and
  Journal of Science}, 2\penalty0 (11):\penalty0 559--572, 1901.

\bibitem[Hotelling(1933)]{hotelling1933analysis}
Harold Hotelling.
\newblock Analysis of a complex of statistical variables into principal
  components.
\newblock \emph{Journal of educational psychology}, 24\penalty0 (6):\penalty0
  417, 1933.

\bibitem[Li and Vit{\'a}nyi(2013)]{li2013introduction}
Ming Li and Paul Vit{\'a}nyi.
\newblock \emph{An introduction to Kolmogorov complexity and its applications}.
\newblock Springer Science \& Business Media, 2013.

\bibitem[Strong et~al.(1998)Strong, Koberle, van Steveninck, and
  Bialek]{strong1998entropy}
Steven~P Strong, Roland Koberle, Rob R de~Ruyter van Steveninck, and William
  Bialek.
\newblock Entropy and information in neural spike trains.
\newblock \emph{Physical review letters}, 80\penalty0 (1):\penalty0 197, 1998.

\bibitem[Paninski(2003)]{paninski2003estimation}
Liam Paninski.
\newblock Estimation of entropy and mutual information.
\newblock \emph{Neural computation}, 15\penalty0 (6):\penalty0 1191--1253,
  2003.

\bibitem[Kolchinsky and Tracey(2017)]{kolchinsky2017estimating}
Artemy Kolchinsky and Brendan Tracey.
\newblock Estimating mixture entropy with pairwise distances.
\newblock \emph{Entropy}, 19\penalty0 (7):\penalty0 361, 2017.

\bibitem[Kraskov et~al.(2004)Kraskov, St{\"o}gbauer, and
  Grassberger]{kraskov2004estimating}
Alexander Kraskov, Harald St{\"o}gbauer, and Peter Grassberger.
\newblock Estimating mutual information.
\newblock \emph{Physical review E}, 69\penalty0 (6):\penalty0 066138, 2004.

\bibitem[Zeng et~al.(2018)Zeng, Xia, and Tong]{zeng2018jackknife}
Xianli Zeng, Yingcun Xia, and Howell Tong.
\newblock Jackknife approach to the estimation of mutual information.
\newblock \emph{Proceedings of the National Academy of Sciences}, 115\penalty0
  (40):\penalty0 9956--9961, 2018.

\bibitem[McAllester and Statos(2018)]{mcallester2018formal}
David McAllester and Karl Statos.
\newblock Formal limitations on the measurement of mutual information.
\newblock \emph{arXiv preprint arXiv:1811.04251}, 2018.

\bibitem[Li and Xie(1996)]{li1996model}
Lei Li and Zhongjie Xie.
\newblock Model selection and order determination for time series by
  information between the past and the future.
\newblock \emph{Journal of time series analysis}, 17\penalty0 (1):\penalty0
  65--84, 1996.

\bibitem[Wiskott and Sejnowski(2002)]{wiskott2002slow}
Laurenz Wiskott and Terrence~J Sejnowski.
\newblock {Slow Feature Analysis}: {Unsupervised} learning of invariances.
\newblock \emph{Neural computation}, 14\penalty0 (4):\penalty0 715--770, 2002.

\bibitem[Bethge et~al.(2007)Bethge, Gerwinn, and Macke]{bethge2007unsupervised}
Matthias Bethge, Sebastian Gerwinn, and Jakob~H Macke.
\newblock Unsupervised learning of a steerable basis for invariant image
  representations.
\newblock In \emph{Human Vision and Electronic Imaging XII}, volume 6492, page
  64920C. International Society for Optics and Photonics, 2007.

\bibitem[Borga(2001)]{borga2001canonical}
Magnus Borga.
\newblock Canonical correlation: a tutorial.
\newblock \emph{On line tutorial http://people. imt. liu. se/magnus/cca},
  4\penalty0 (5), 2001.

\bibitem[Cunningham and Byron(2014)]{cunningham2014dimensionality}
John~P Cunningham and M~Yu Byron.
\newblock Dimensionality reduction for large-scale neural recordings.
\newblock \emph{Nature neuroscience}, 17\penalty0 (11):\penalty0 1500, 2014.

\bibitem[Tegmark(2019)]{tegmark2019optimal}
Max Tegmark.
\newblock Optimal latent representations: Distilling mutual information into
  principal pairs.
\newblock \emph{arXiv preprint arXiv:1902.03364}, 2019.

\bibitem[Larsen(2002)]{larsen2002decomposition}
Rasmus Larsen.
\newblock Decomposition using {Maximum Autocorrelation Factors}.
\newblock \emph{Journal of Chemometrics: A Journal of the Chemometrics
  Society}, 16\penalty0 (8-10):\penalty0 427--435, 2002.

\bibitem[Hyv{\"a}rinen(2001)]{hyvarinen2001complexity}
Aapo Hyv{\"a}rinen.
\newblock Complexity pursuit: separating interesting components from time
  series.
\newblock \emph{Neural computation}, 13\penalty0 (4):\penalty0 883--898, 2001.

\bibitem[Goerg(2013)]{goerg2013forecastable}
Georg Goerg.
\newblock Forecastable component analysis.
\newblock In \emph{International Conference on Machine Learning}, pages 64--72,
  2013.

\bibitem[Tong et~al.(1990)Tong, Soon, Huang, and Liu]{tong1990amuse}
Lang Tong, VC~Soon, Yih-Fang Huang, and RALR Liu.
\newblock {AMUSE}: a new blind identification algorithm.
\newblock In \emph{IEEE international symposium on circuits and systems}, pages
  1784--1787. IEEE, 1990.

\bibitem[Ziehe and M{\"u}ller(1998)]{ziehe1998tdsep}
Andreas Ziehe and Klaus-Robert M{\"u}ller.
\newblock {TDSEP}—an efficient algorithm for blind separation using time
  structure.
\newblock In \emph{International Conference on Artificial Neural Networks},
  pages 675--680. Springer, 1998.

\bibitem[St{\"o}gbauer et~al.(2004)St{\"o}gbauer, Kraskov, Astakhov, and
  Grassberger]{stogbauer2004least}
Harald St{\"o}gbauer, Alexander Kraskov, Sergey~A Astakhov, and Peter
  Grassberger.
\newblock Least-dependent-component analysis based on mutual information.
\newblock \emph{Physical Review E}, 70\penalty0 (6):\penalty0 066123, 2004.

\bibitem[Richthofer and Wiskott(2015)]{richthofer2015predictable}
Stefan Richthofer and Laurenz Wiskott.
\newblock {Predictable Feature Analysis}.
\newblock In \emph{2015 IEEE 14th International Conference on Machine Learning
  and Applications (ICMLA)}, pages 190--196. IEEE, 2015.

\bibitem[Weghenkel et~al.(2017)Weghenkel, Fischer, and
  Wiskott]{weghenkel2017graph}
Bj{\"o}rn Weghenkel, Asja Fischer, and Laurenz Wiskott.
\newblock {Graph-based Predictable Feature Analysis}.
\newblock \emph{Machine Learning}, 106\penalty0 (9-10):\penalty0 1359--1380,
  2017.

\bibitem[Creutzig et~al.(2009)Creutzig, Globerson, and
  Tishby]{creutzig2009past}
Felix Creutzig, Amir Globerson, and Naftali Tishby.
\newblock Past-future information bottleneck in dynamical systems.
\newblock \emph{Physical Review E}, 79\penalty0 (4):\penalty0 041925, 2009.

\bibitem[Thirion and Faugeras(2003)]{thirion2003dynamical}
Bertrand Thirion and Olivier Faugeras.
\newblock {Dynamical components analysis of fMRI data through kernel PCA}.
\newblock \emph{NeuroImage}, 20\penalty0 (1):\penalty0 34--49, 2003.

\bibitem[Seifert et~al.(2018)Seifert, Korn, Hartmann, and
  Uhl]{seifert2018dynamical}
Bastian Seifert, Katharina Korn, Steffen Hartmann, and Christian Uhl.
\newblock {Dynamical Component Analysis (DyCA): dimensionality reduction for
  high-dimensional deterministic time-series}.
\newblock In \emph{2018 IEEE 28th International Workshop on Machine Learning
  for Signal Processing (MLSP)}, pages 1--6. IEEE, 2018.

\bibitem[Korn et~al.(2019)Korn, Seifert, and Uhl]{korn2019dynamical}
Katharina Korn, Bastian Seifert, and Christian Uhl.
\newblock {Dynamical Component Analysis (DyCA) and its application on epileptic
  EEG}.
\newblock In \emph{ICASSP 2019-2019 IEEE International Conference on Acoustics,
  Speech and Signal Processing (ICASSP)}, pages 1100--1104. IEEE, 2019.

\bibitem[Kalman(1960)]{kalman1960new}
Rudolph~Emil Kalman.
\newblock A new approach to linear filtering and prediction problems.
\newblock \emph{Journal of basic Engineering}, 82\penalty0 (1):\penalty0
  35--45, 1960.

\bibitem[Yu et~al.(2009)Yu, Cunningham, Santhanam, Ryu, Shenoy, and
  Sahani]{byron2009gaussian}
Byron~M Yu, John~P Cunningham, Gopal Santhanam, Stephen~I Ryu, Krishna~V
  Shenoy, and Maneesh Sahani.
\newblock Gaussian-process factor analysis for low-dimensional single-trial
  analysis of neural population activity.
\newblock \emph{J Neurophysiol}, 102:\penalty0 614--635, 2009.

\bibitem[Pandarinath et~al.(2018)Pandarinath, O’Shea, Collins, Jozefowicz,
  Stavisky, Kao, Trautmann, Kaufman, Ryu, Hochberg,
  et~al.]{pandarinath2018inferring}
Chethan Pandarinath, Daniel~J O’Shea, Jasmine Collins, Rafal Jozefowicz,
  Sergey~D Stavisky, Jonathan~C Kao, Eric~M Trautmann, Matthew~T Kaufman,
  Stephen~I Ryu, Leigh~R Hochberg, et~al.
\newblock Inferring single-trial neural population dynamics using sequential
  auto-encoders.
\newblock \emph{Nature methods}, page~1, 2018.

\bibitem[O'Doherty et~al.(2017)O'Doherty, Cardoso, Makin, and
  Sabes]{o_doherty_joseph_e_2017_583331}
Joseph~E. O'Doherty, Mariana M.~B. Cardoso, Joseph~G. Makin, and Philip~N.
  Sabes.
\newblock {Nonhuman Primate Reaching with Multichannel Sensorimotor Cortex
  Electrophysiology}, May 2017.
\newblock URL \url{https://doi.org/10.5281/zenodo.583331}.

\bibitem[Mizuseki et~al.(2009)Mizuseki, Sirota, Pastalkova, and
  Buzs{\'a}ki]{mizuseki2009multi}
K~Mizuseki, A~Sirota, E~Pastalkova, and G~Buzs{\'a}ki.
\newblock Multi-unit recordings from the rat hippocampus made during open field
  foraging.
\newblock \emph{Available online at: CRCNS. org}, 2009.

\bibitem[Glaser et~al.(2017)Glaser, Chowdhury, Perich, Miller, and
  Kording]{glaser2017machine}
Joshua~I Glaser, Raeed~H Chowdhury, Matthew~G Perich, Lee~E Miller, and
  Konrad~P Kording.
\newblock Machine learning for neural decoding.
\newblock \emph{arXiv preprint arXiv:1708.00909}, 2017.

\bibitem[Gene(2017)]{gene2017}
Selfish Gene.
\newblock Historical hourly weather data 2012-2017, Dec 2017.
\newblock URL
  \url{https://www.kaggle.com/selfishgene/historical-hourly-weather-data}.

\bibitem[Malekzadeh et~al.(2018)Malekzadeh, Clegg, Cavallaro, and
  Haddadi]{malekzadeh2018protecting}
Mohammad Malekzadeh, Richard~G Clegg, Andrea Cavallaro, and Hamed Haddadi.
\newblock Protecting sensory data against sensitive inferences.
\newblock In \emph{Proceedings of the 1st Workshop on Privacy by Design in
  Distributed Systems}, page~2. ACM, 2018.

\bibitem[Churchland et~al.(2012)Churchland, Cunningham, Kaufman, Foster,
  Nuyujukian, Ryu, and Shenoy]{churchland2012neural}
Mark~M Churchland, John~P Cunningham, Matthew~T Kaufman, Justin~D Foster, Paul
  Nuyujukian, Stephen~I Ryu, and Krishna~V Shenoy.
\newblock Neural population dynamics during reaching.
\newblock \emph{Nature}, 487\penalty0 (7405):\penalty0 51, 2012.

\bibitem[Wilson and McNaughton(1993)]{wilson1993dynamics}
Matthew~A Wilson and Bruce~L McNaughton.
\newblock Dynamics of the hippocampal ensemble code for space.
\newblock \emph{Science}, 261\penalty0 (5124):\penalty0 1055--1058, 1993.

\bibitem[Golub et~al.(2018)Golub, Sadtler, Oby, Quick, Ryu, Tyler-Kabara,
  Batista, Chase, and Yu]{golub2018learning}
Matthew~D Golub, Patrick~T Sadtler, Emily~R Oby, Kristin~M Quick, Stephen~I
  Ryu, Elizabeth~C Tyler-Kabara, Aaron~P Batista, Steven~M Chase, and Byron~M
  Yu.
\newblock Learning by neural reassociation.
\newblock \emph{Nature neuroscience}, 21\penalty0 (4):\penalty0 607--616, 2018.

\bibitem[Costa et~al.(2019)Costa, Ahamed, and Stephens]{costa2019adaptive}
Antonio~C Costa, Tosif Ahamed, and Greg~J Stephens.
\newblock Adaptive, locally linear models of complex dynamics.
\newblock \emph{Proceedings of the National Academy of Sciences}, 116\penalty0
  (5):\penalty0 1501--1510, 2019.

\bibitem[Tibshirani(1996)]{tibshirani1996regression}
Robert Tibshirani.
\newblock Regression shrinkage and selection via the lasso.
\newblock \emph{Journal of the Royal Statistical Society: Series B
  (Methodological)}, 58\penalty0 (1):\penalty0 267--288, 1996.

\bibitem[Paszke et~al.(2017)Paszke, Gross, Chintala, and
  Chanan]{paszke2017pytorch}
Adam Paszke, Sam Gross, Soumith Chintala, and Gregory Chanan.
\newblock {PyTorch}: {Tensors} and dynamic neural networks in {Python} with
  strong {GPU} acceleration.
\newblock \emph{{PyTorch}: {Tensors} and dynamic neural networks in {Python}
  with strong {GPU} acceleration}, 6, 2017.

\bibitem[Zhu et~al.(1997)Zhu, Byrd, Lu, and Nocedal]{zhu1997algorithm}
Ciyou Zhu, Richard~H Byrd, Peihuang Lu, and Jorge Nocedal.
\newblock {Algorithm 778: L-BFGS-B: Fortran subroutines for large-scale
  bound-constrained optimization}.
\newblock \emph{ACM Transactions on Mathematical Software (TOMS)}, 23\penalty0
  (4):\penalty0 550--560, 1997.

\bibitem[Jones et~al.(2001--)Jones, Oliphant, Peterson, et~al.]{jones2001scipy}
Eric Jones, Travis Oliphant, Pearu Peterson, et~al.
\newblock {SciPy}: Open source scientific tools for {Python}, 2001--.
\newblock URL \url{http://www.scipy.org/}.
\newblock [Online; accessed <today>].

\bibitem[Strogatz(2018)]{strogatz2018nonlinear}
Steven~H Strogatz.
\newblock \emph{Nonlinear Dynamics and Chaos with Student Solutions Manual:
  With Applications to Physics, Biology, Chemistry, and Engineering}.
\newblock CRC Press, 2018.

\bibitem[Li(2006)]{li2006some}
Lei~M Li.
\newblock Some notes on mutual information between past and future.
\newblock \emph{Journal of Time Series Analysis}, 27\penalty0 (2):\penalty0
  309--322, 2006.

\bibitem[Creutzig and Sprekeler(2008)]{creutzig2008predictive}
Felix Creutzig and Henning Sprekeler.
\newblock Predictive coding and the slowness principle: {An}
  information-theoretic approach.
\newblock \emph{Neural Computation}, 20\penalty0 (4):\penalty0 1026--1041,
  2008.

\bibitem[Amir et~al.(2015)Amir, Tiomkin, and Tishby]{amir2015past}
Nadav Amir, Stas Tiomkin, and Naftali Tishby.
\newblock Past-future information bottleneck for linear feedback systems.
\newblock In \emph{2015 54th IEEE Conference on Decision and Control (CDC)},
  pages 5737--5742. IEEE, 2015.

\bibitem[Chechik et~al.(2005)Chechik, Globerson, Tishby, and
  Weiss]{chechik2005information}
Gal Chechik, Amir Globerson, Naftali Tishby, and Yair Weiss.
\newblock Information bottleneck for {Gaussian} variables.
\newblock \emph{Journal of machine learning research}, 6\penalty0
  (Jan):\penalty0 165--188, 2005.

\bibitem[Ghahramani and Hinton(1996)]{ghahramani1996parameter}
Zoubin Ghahramani and Geoffrey~E Hinton.
\newblock Parameter estimation for linear dynamical systems.
\newblock Technical report, Technical Report CRG-TR-96-2, University of
  Totronto, Dept. of Computer Science, 1996.

\end{thebibliography}

\end{document}